\documentstyle[12pt]{article}
\def\journal{\topmargin .3in	\oddsidemargin .5in
	\headheight 0pt	\headsep 0pt
	\textwidth 5.625in 
	\textheight 8.25in 
	\marginparwidth 1.5in
	\parindent 2em
	\parskip .5ex plus .1ex		\jot = 1.5ex}
\journal

\catcode`\@=11
\def\marginnote#1{}
\catcode`\@=11
\def\section{\@startsection {section}{1}{0pt}{-3.5ex plus -1ex minus
 -.2ex}{2.3ex plus .2ex}{\raggedright\large\bf}}
\catcode`\@=12
\newskip\humongous \humongous=0pt plus 1000pt minus 1000pt
\def\caja{\mathsurround=0pt}
\def\eqalign#1{\,\vcenter{\openup1\jot \caja
	\ialign{\strut \hfil$\displaystyle{##}$&$
	\displaystyle{{}##}$\hfil\crcr#1\crcr}}\,}
\newif\ifdtup
\def\panorama{\global\dtuptrue \openup1\jot \caja
	\everycr{\noalign{\ifdtup \global\dtupfalse
	\vskip-\lineskiplimit \vskip\normallineskiplimit
	\else \penalty\interdisplaylinepenalty \fi}}}
\def\eqalignno#1{\panorama \tabskip=\humongous
	\halign to\displaywidth{\hfil$\displaystyle{##}$
	\tabskip=0pt&$\displaystyle{{}##}$\hfil
	\tabskip=\humongous&\llap{$##$}\tabskip=0pt
	\crcr#1\crcr}}
\def\R{{\rm I\!R}}
\def\N{{\rm I\!N}}
\def\one{{\mathchoice {\rm 1\mskip-4mu l} {\rm 1\mskip-4mu}
{\rm 1\mskip-4.5mu l} {\rm 1\mskip-5mu l}}}
\def\Q{{\mathchoice
{\setbox0=\hbox{$\displaystyle\rm Q$}\hbox{\raise 0.15\ht0\hbox to0pt
{\kern0.4\wd0\vrule height0.8\ht0\hss}\box0}}
{\setbox0=\hbox{$\textstyle\rm Q$}\hbox{\raise 0.15\ht0\hbox to0pt
{\kern0.4\wd0\vrule height0.8\ht0\hss}\box0}}
{\setbox0=\hbox{$\scriptstyle\rm Q$}\hbox{\raise 0.15\ht0\hbox to0pt
{\kern0.4\wd0\vrule height0.7\ht0\hss}\box0}}
{\setbox0=\hbox{$\scriptscriptstyle\rm Q$}\hbox{\raise 0.15\ht0\hbox to0pt
{\kern0.4\wd0\vrule height0.7\ht0\hss}\box0}}}}
\def\C{{\mathchoice
{\setbox0=\hbox{$\displaystyle\rm C$}\hbox{\hbox to0pt
{\kern0.4\wd0\vrule height0.9\ht0\hss}\box0}}
{\setbox0=\hbox{$\textstyle\rm C$}\hbox{\hbox to0pt
{\kern0.4\wd0\vrule height0.9\ht0\hss}\box0}}
{\setbox0=\hbox{$\scriptstyle\rm C$}\hbox{\hbox to0pt
{\kern0.4\wd0\vrule height0.9\ht0\hss}\box0}}
{\setbox0=\hbox{$\scriptscriptstyle\rm C$}\hbox{\hbox to0pt
{\kern0.4\wd0\vrule height0.9\ht0\hss}\box0}}}}

\font\fivesans=cmss10 at 4.61pt
\font\sevensans=cmss10 at 6.81pt
\font\tensans=cmss10
\newfam\sansfam
\textfont\sansfam=\tensans\scriptfont\sansfam=\sevensans\scriptscriptfont
\sansfam=\fivesans
\def\sans{\fam\sansfam\tensans}
\def\Z{{\mathchoice
{\hbox{$\sans\textstyle Z\kern-0.4em Z$}}
{\hbox{$\sans\textstyle Z\kern-0.4em Z$}}
{\hbox{$\sans\scriptstyle Z\kern-0.3em Z$}}
{\hbox{$\sans\scriptscriptstyle Z\kern-0.2em Z$}}}}

\mathchardef\endbar="375

\def\ceilfill{$\raise3pt\hbox{$\mathsurround=0pt\mathord\endbar$}
  \mkern-2mu \xleaders\hbox{$\mkern-5mu
  \mathord-\mkern-5mu$}\hfill\mkern-7mu
  \raise3pt\hbox{$\mathsurround=0pt\mathord\endbar$}$}

\def\floorfill{$\raise9pt\hbox{$\mathsurround=0pt\mathord\endbar$}
  \mkern-2mu \xleaders\hbox{$\mkern-5mu
  \mathord-\mkern-5mu$}\hfill\mkern-7mu
  \raise9pt\hbox{$\mathsurround=0pt\mathord\endbar$}$}

\def\overcontract#1{\mathop{\vbox{\ialign{##\crcr\noalign{\kern3pt}
  \ceilfill\hskip6pt\crcr\noalign{\kern3pt\nointerlineskip}
  $\hfil\displaystyle{#1}\hfil$\crcr}}}}

\def\undercontract#1{\mathop{\vtop{\ialign{##\crcr
  $\hfil\displaystyle{#1}\hfil$\crcr\noalign{\kern3pt\nointerlineskip}
  \floorfill\hskip6pt\crcr\noalign{\kern3pt}}}}}
\def\a{\alpha}
\def\b{\beta}
\def\g{\gamma}
\def\d{\delta}

\def\m{\mu}
\def\n{\nu}

\def\p{\pi}
\def\ps{\psi}

\def\l{\lambda}
\def\o{\omega}
\def\f{\phi}
\def\x{\xi}
\def\et{\eta}

\def\Ga{\Gamma}

\def\O{\Omega}
\def\D{\Delta}
\def\bz{\bar{z}}
\def\bc{\bar{\g}}

\def\bu{\bar{u}}

\def\br{\bar{r}}

\def\fb{\bar{f}}

\def\tf{\bar{\phi}}
\def\bps{\bar{\psi}}

\def\bM{\bar{M}}
\def\bA{\bar{A}}

\def\bY{\bar{Y}}

\def\bX{\bar{X}}

\def\tL{\tilde{L}}
\def\tT{\tilde{T}}
\def\tD{\tilde{\Delta}}

\def\F{{\cal F}}

\def\P{{\cal P}}

\def\cD{{\cal D}}
\def\T{{\cal T}}
\def\C{{\cal C}}
\def\cO{{\cal O}}
\def\Y{{\cal Y}}
\def\N{{\cal N}}

\def\pa{\partial}
\def\bpa{\bar{\partial}}
\def\ve{\vert}

\def\ra{\rightarrow}
\def\lra{\leftrightarrow}

\def\oti{\otimes}

\def\xx{\hbox{ }^*_*}
\def\ch#1#2{\left( #1 \atop #2 \right)}
\def\cl#1#2#3#4#5#6{\left( \matrix{ #1 & #2 & #3 \cr #4 & #5 & #6 \cr} \right)}

\def\ref#1{$^{#1)}$}

\begin{document}
\begin{titlepage}
\begin{center}
 June 1993        \hfill     UCB-PTH-93/18 \\
hep-th/9305072    \hfill     LBL-34111 \\
revised version   \hfill    BONN-HE-93/17
\vskip .05in

{\large \bf Solving the Ward Identities \\
 of Irrational Conformal Field Theory }

\vskip .15in
M.B. Halpern
\footnote{e-mail: MBHALPERN@LBL.GOV, THEORY::HALPERN}
\vskip .05in
{\em  Department of Physics, University of California\\
      Theoretical Physics Group, Lawrence Berkeley Laboratory\\
      Berkeley, California 94720 \\
       USA}
\vskip .03in
\vskip .03in
N.A. Obers
\footnote{e-mail: OBERS@PIB1.PHYSIK.UNI-BONN.DE, 13581::OBERS}
\vskip .03in
{\em Physikalisches Institut der Universit\"at Bonn \\
Nu{\ss}allee 12, D-53115 Bonn  \\
Germany}

\end{center}

\begin{abstract}
The affine-Virasoro Ward identities are a system of non-linear differential
equations which describe the correlators of all affine-Virasoro constructions,
including rational and irrational conformal field theory. We study the Ward
identities in some detail, with several central results. First, we solve for
the correlators of the affine-Sugawara nests, which are associated to the
nested subgroups $g\supset h_1 \supset \ldots \supset h_n$. We also find an
equivalent algebraic formulation which allows us to find global solutions
across the set of all affine-Virasoro constructions. A particular global
solution is discussed which gives the correct nest correlators, exhibits
braiding for all affine-Virasoro correlators, and shows good physical behavior,
at least for four-point correlators at high level on simple $g$. In rational
and irrational conformal field theory, the high-level fusion rules of the
broken affine modules follow the Clebsch-Gordan coefficients of the
representations.
\end{abstract}

\end{titlepage}
\renewcommand{\thepage}{\arabic{page}}
\setcounter{page}{1}
\setcounter{footnote}{1}
\section{Introduction}
Affine Lie algebra [1,2], or current algebra on $S^1$, was discovered
independently in mathematics and physics. Affine-Virasoro constructions are
the most general Virasoro operators [3,4]
$$T =L^{ab} J_a J_b \eqno(1.1) $$
which are quadratic in the currents $J_a$ of the affine algebra. The
coefficients $L^{ab}$ in the stress tensor $T$ can be any solution of the
Virasoro master equation. The solution space of the master equation, called
affine-Virasoro space, includes the affine-Sugawara constructions
[2,5,6,7], the coset constructions [2,5,8], the
affine-Sugawara nests [9,10,11], and a vast number of new
constructions, most of which have irrational central charge [10]. As an
example, it is known that there are  approximately $1/4$ billion solutions of
the master equation on each level of affine $SU(3)$, while the value at level
5 [12]
$$ c \left( (SU(3)_5)_{D(1)}^{\#} \right) = 2 \left( 1 - {1 \over \sqrt{61}}
\right) \simeq 1.7439 \eqno(1.2) $$
is the lowest unitary irrational central charge yet observed. Partial
classification of the solution space and other developments in the Virasoro
master equation are reviewed in Ref.[13].

It is clear that the Virasoro master equation is the first step in the study of
irrational conformal field theory (ICFT), which includes rational conformal
field theory (RCFT) as a small subspace of relatively high symmetry,
$$ {\rm ICFT} \supset {\rm RCFT} \;\;\;\;. \eqno(1.3) $$
The next step is a description of the correlators of irrational conformal field
theory, about which we have the intuitive notion that they must involve a
generically-infinite number of conformal structures. The organization of these
structures must be generically new, since it is unlikely that the generic
affine-Virasoro construction supports an extended chiral algebra.

Using null states of the Knizhnik-Zamolodchikov type [7], we recently reported
the derivation of dynamical equations, the factorized affine-Virasoro Ward
identities [14], which describe the correlators of irrational conformal
field theory. For any K-conjugate pair [2,5,8,3] of affine-Virasoro
constructions, one may compute a set of affine-Virasoro connections for the
pair. The connections are the input data for the
Ward identities, which generalize the Knizhnik-Zamolodchikov equations to the
broader context of coset constructions  and irrational conformal field theory.
In form, the Ward identities are a system of coupled non-linear differential
equations for the factorized correlators of the K-conjugate pair of conformal
field theories. Because the affine-Sugawara constructions are K-conjugate to
the trivial theory, the Knizhnik-Zamolodchikov equations are included as the
simplest case of the Ward identities.

As a first non-trivial example, we solved the Ward identities for the
correlators of the coset constructions, providing a derivation  of the coset
blocks of Douglas [15]. In the present paper, we go beyond the cosets
to study the Ward identities for all affine-Virasoro constructions.

After a brief review of the master equation and the Ward identities, we discuss
some general properties of the affine-Virasoro connections, including
K-conjugation covariance, crossing symmetry and the high-level form of the
general connections. As a first application of the general properties, we
construct the connections and solve for the conformal correlators of all the
affine-Sugawara nests, which are associated to the  nested subgroups
$$g \supset h_1 \supset \ldots  \supset h_n \;\;\;\;. \eqno(1.4)$$
Our results explicitly verify the intuition that the nests are tensor-product
RCFT's, constructed by tensoring the relevant coset and subgroup constructions.

Turning to the general affine-Virasoro construction, we find an equivalent
algebraic formulation of the system which, given the affine-Virasoro
connections, allows us to find global solutions of the Ward identities across
all affine-Virasoro space. The solutions involve a generically-infinite number
of conformal structures, in accord with intuitive notions about irrational
conformal field theory, but many of the solutions are apparently not physical.

Based on a natural eigenvalue problem in the system, we focus on a particular
infinite-dimensional global solution with the following properties:
\begin{enumerate}
\item[a)] The conformal structures are degenerate for the coset constructions
and affine-Sugawara nests, and the correct correlators are obtained in these
cases.
\item[b)] The solution exhibits a braiding for all affine-Virasoro correlators
which includes and generalizes the braiding of rational conformal field theory.
The origin of the braiding is the linearity of the eigenvalue problem.
\item[c)] The solution shows good physical behavior, at least for four-point
affine-Virasoro correlators at high level on simple $g$. From the high-level
correlators, we determine that the high-level fusion rules of the broken
affine modules follow the Clebsch-Gordan coefficients of the representations.
\end{enumerate}

\section{The Virasoro Master Equation}
In this section, we review the Virasoro master equation and some features
of the system which will be useful below.

The general affine-Virasoro construction begins with the currents $J_a$ of
untwisted affine $g$ [1,2]
$$ J_a(z) = \sum_m J_a^{(m)} z^{-m-1}\;\;\;\;,\;\;\;a =1, \ldots , {\rm dim}\,g
\;\; \;, \;\;\;\;m,n \in \Z \eqno(2.1a) $$
$$J_a(z)\, J_b(w)={G_{ab} \over (z-w)^2}+ { i{f_{ab}}^c \over z-w} \, J_c(w) +
\cO (z-w)^0 \eqno(2.1b) $$
where ${f_{ab}}^c$ and $G_{ab}$ are the structure constants and general Killing
metric of $g$. The current algebra (2.1) is completely general since $g$ is not
necessarily compact or semisimple. In particular, to obtain level
$x_I = 2k_I/\psi_I^2$ of $g_I$ in $g = \oplus_I g_I$ with dual Coxeter number
$\tilde{h}_I=Q_I/\psi_I^2$, take
$$G_{ab}=\oplus_I k_I \, \et_{ab}^I\;\;\;\;,\;\;\;\;
{f_{ac}}^d {f_{bd}}^c = - \oplus_I Q_I \, \et_{ab}^I \eqno(2.2) $$
where  $\et_{ab}^I$ and $\psi_I$ are the Killing metric and the highest root of
$g_I$.

Next, consider the set of operators quadratic in the currents
$$ T(z)= L^{ab} \xx J_a (z)J_b (z)\xx =\sum_m L^{(m)}z^{-m-2} \eqno(2.3) $$
where the set of coefficients $L^{ab}= L^{ba}$ is called the inverse inertia
tensor, in analogy with the spinning top. The requirement that $T(z)$ is a
Virasoro operator
$$T(z)\,T(w)={c/2 \over (z-w)^4} +\left(\frac{2}{(z-w)^2}+{ \pa_w \over z-w}
\right)T(w)+ \cO (z-w)^0 \eqno(2.4)$$
restricts the values of the inverse inertia tensor to those which solve the
Virasoro master equation [3,4]
$$ L^{ab}=2 L^{ac}G_{cd}L^{db}-L^{cd}L^{ef}{f_{ce}}^{a}{f_{df}}^{b}-
L^{cd}{f_{ce}}^{f}{f_{df}}^{(a}L^{b)e} \eqno(2.5a)$$
$$ c=2 \,G_{ab}L^{ab}\;\;\;\;.\eqno(2.5b)$$
The Virasoro master equation has been identified [16] as an Einstein-like
system on the group manifold, with $L^{ab}$ the inverse metric on tangent space
and $c={\rm dim}\,g-4R$, where $R$ is the Einstein curvature scalar.

Some general features of the Virasoro master equation include:

\noindent 1. Affine-Sugawara constructions [2,5,6,7]. The
affine-Sugawara (A-S) construction $L_g$ is
$$ L_g^{ab}=\oplus_I {\eta_I^{ab} \over 2k_I +Q_I}\;\;\;\,,\;\;\;\;\;\;\;
c_g=\sum_I {x_I \, {\rm dim}\,g_I \over x_I + \tilde{h}_I } \eqno(2.6) $$
for arbitrary level of any $g$, and similarly for $L_h$ when $h \subset g$.

\noindent 2. K-conjugation covariance [2,5,8,3]. When $L$ is a
solution of the master equation on $g$, then so is the K-conjugate partner
$\tilde{L}$ of $L$,
$$ \tilde{L}^{ab}=L_g^{ab}-L^{ab}\;\;\;,\;\;\;\;\;\;\;\tilde{c}=c_g-c
\eqno(2.7) $$
and the corresponding stress tensors form a K-conjugate pair of commuting
Virasoro operators
$$ \tilde{T}(z) =\sum_m \tilde{L}^{(m)} z^{-m-2} \eqno(2.8a) $$
$$\tilde{T}(z)\,\tilde{T}(w)={\tilde{c}/2 \over
(z-w)^4}+\left(\frac{2}{(z-w)^2}
+\frac{1}{z-w} \partial_{w}\right)\tilde{T}(w)+ \cO (z-w)^0 \eqno(2.8b)$$
$$ T(z)\, \tilde{T} (w)= \cO (z-w)^0  \;\;\;\;\;\; . \eqno(2.8c) $$
The affine-Virasoro stress tensors $T$ and $\tT$ are quasi (2,0) operators
under the affine-Sugawara stress tensor $T_g=T + \tT$.

\noindent 3. Cosets and affine-Sugawara nests. The simplest K-conjugate pairs
are the subgroup constructions $L_h$ and the corresponding $g/h$ coset
constructions [2,5,8]
$$ L_{g/h}^{ab} = L_g^{ab} - L_h^{ab}\;\;\;,\;\;\;\;c_{g/h}=c_g-c_h\eqno(2.9)$$
while repeated K-conjugation on  subgroup sequences
$g \supset h_1 \supset \ldots \supset h_n$ generates the affine-Sugawara
(A-S) nests [9,10,11]
$$ \eqalignno{
L_{g/h_1 / \ldots /h_n} & = L_g - L_{h_1/ \ldots /h_n} = L_g + \sum_{j=1}^n
(-)^j
L_{h_j}  &(2.10a) \cr
 c_{g/h_1/ \ldots / h_n} & =c_g - c_{h_1/\ldots /h_n} =c_g + \sum_{j=1}^n (-)^j
c_{h_j} \;\;\;\;.  &(2.10b) \cr} $$
Note that the stress tensors of the affine-Sugawara nests may be written as
sums of mutually-commuting Virasoro constructions on $g/h$ and $h$
$$ \eqalignno{
 T_{g/h_1/ \ldots  /h_{2n+1} } & = T_{g/h_1} + \sum_{i=1}^{n}
T_{h_{2i}/h_{2i+1}}  &(2.11a) \cr
 T_{g/h_1/ \ldots  /h_{2n} } & = T_{g/h_1} + \sum_{i=1}^{n-1}
T_{h_{2i}/h_{2i+1}} + T_{h_{2n}} &(2.11b) \cr} $$
so the conformal field theories of the affine-Sugawara nests are expected to
be tensor-product theories. By computation of the nest correlators, we will
explicitly verify this intuition in Section 5.

Other developments in the Virasoro master equation, including the worldsheet
action [17] for the generic affine-Virasoro construction, are reviewed in
Ref.[13]

\section{The Affine-Virasoro Ward Identities}
In this section, we review the affine-Virasoro (A-V) Ward identities [14],
which generalize the Knizhnik-Zamolodchikov equations [7] to the broader
context of coset constructions and irrational conformal field theory.

\noindent 1. Virasoro biprimary states. Let $| 0 \rangle$ be the affine vacuum
and $R_g(\T,z)$  be the affine primary fields corresponding to irreducible
matrix representation $\T$ of $g$. Under the stress tensor $T$, the conformal
weights $\D_{\a}(\T)$ of the affine primary states are the eigenvalues of the
conformal weight matrix $L^{ab} \T_a \T_b $ [18,10]. In what follows, we
choose an $L$-basis of $\T$  [14], in which the conformal weight matrix is
diagonal
$$  L^{ab} {(\T_a \T_b)_{\a}}^{\b} = \D_{\a}(\T) \, \d_{\a}^{\b}
\;\;\;\;,\;\;\;\; \a,\b =1 , \ldots ,{\rm dim}\,\T \;\;\;\; . \eqno(3.1) $$
Then the corresponding eigenstates $R_g^{\a}(\T,0) | 0 \rangle$, called the
$L^{ab}$-broken affine primary states, satisfy
$$ L^{m \geq 0} R_g^{\a} (\T,0) |0\rangle =\d_{m,0} \, \D_{\a}(\T)
R_g^{\a} (\T,0) |0\rangle \eqno(3.2a) $$
$$ \tilde{L}^{m \geq 0} R_g^{\a} (\T,0) |0\rangle =\d_{m,0} \,
\tilde{\D}_{\a}(\T)  R_g^{\a} (\T,0) |0\rangle \eqno(3.2b) $$
$$  \D_{g}(\T) =  \D_{\a}(\T) + \tilde{\D}_{\a}(\T)  \eqno(3.2c) $$
where $\D_g (\T)$ and $\tD_{\a} (\T)$ are the conformal weights of $T_g$ and
$\tT$ respectively. The $L^{ab}$-broken affine primary states are examples of
Virasoro biprimary states [18,10,14], which are simultaneously Virasoro
primary under the K-conjugate stress tensors $T$ and $\tT$.

\noindent 2. Virasoro biprimary fields. The corresponding $L^{ab}$-broken
affine primary fields
$$ R^{\a}(\T,\bz,z) ={\rm e}^{(\bz-z) \tL^{(-1)} } R_g^{\a} (\T,z) \,
{\rm e}^{(z-\bz) \tL^{(-1)}} = {\rm e}^{(z-\bz) L^{(-1)} } R_g^{\a} (\T,\bz) \,
{\rm e}^{(\bz-z) L^{(-1)}} \eqno (3.3a) $$
$$ R^{\a} (\T,z,z) = R_g^{\a} (\T,z) \;\;\;\;,\;\;\;\;
 R^{\a}(\T,0,0)  |0 \rangle= R_g^{\a} (\T,0) |0 \rangle  \eqno(3.3b) $$
are examples of Virasoro biprimary fields [18,14], which are
simultaneously Virasoro primary under $T$ and $\tT$. The correlators of
Virasoro biprimary fields, such as
$$ A^{\a}(\bz,z) =  \langle R^{\a_1}(\T^1,\bz_1,z_1) \ldots
R^{\a_n}(\T^n,\bz_n,z_n) \rangle \;\;\;\;,\;\;\;\;\a=(\a_1\ldots\a_n)
\eqno(3.4) $$
are called biconformal correlators.

\noindent 3. Affine-Virasoro Ward identities [14]. Using null states of
the Knizhnik-Zamolodchikov type, one obtains the affine-Virasoro Ward
identities
for the biconformal correlators. In the case of broken affine primary fields,
these read
$$   \bpa_{j_1} \ldots \bpa_{j_q} \pa_{i_1} \ldots \pa_{i_p} A^{\a}(\bz,z)
 \ve_{\bz=z} = A^{\b}_g (z) { W_{j_1 \ldots j_q, i_1 \ldots i_p }
(z)_{\b}}^{\a} \eqno(3.5a) $$
$$ A_g^{\a}(z)  =A^{\a} (z,z)=  \langle
R_g^{\a_1}(\T^1,z_1) \ldots R_g^{\a_n}(\T^n,z_n) \rangle \eqno(3.5b) $$
where $W_{ j_1 \ldots j_q ,i_1 \ldots i_p} $ are the affine-Virasoro
connections and $A_g(z)$ is the affine-Sugawara correlator, which solves
the Knizhnik-Zamolodchikov equations [7] and the $g$-global Ward identities
$$ A_g^{\b} ( \sum_{i=1}^n \T_a^i)_{\b}{}^{\a} =0 \;\;\;\;,\;\;\;\;a \in g
\;\;\;\;. \eqno(3.6) $$
The connections may be computed by
standard dispersive techniques from the formula
$$  A_g^{\b} (z) {W_{j_1 \ldots j_q, i_1 \ldots i_p } (z)_{\b}}^{\a} =
\;\; \;\;\;\;\;\;\;\;\;\;\; \;\;\;\;\;\;\;\;\;\; $$
$$ \left[ \prod_{r=1}^{q} \tL^{a_r b_r} \oint_{z_{j_r}}
{{\rm d}\o_r \over 2\p i} \oint_{\o_r}  {{\rm d}\et_r \over 2\p i} \;
{1 \over \et_r-\o_r} \right] \!
\left[ \prod_{s=1}^{p} L^{c_s d_s} \oint_{z_{i_s}} \!
{{\rm d}\o_{q+s} \over 2\p i} \oint_{\o_{q+s}} \!
{{\rm d}\et_{q+s} \over 2\p i} \; {1 \over \et_{q+s}-\o_{q+s}} \right] $$
$$ \times  \langle
J_{a_1}(\et_1)  J_{b_1}(\o_1)  \ldots J_{a_q}(\et_q)  J_{b_q}(\o_q)
J_{c_1}(\et_{q+1})  J_{d_1}(\o_{q+1})  \ldots  $$
$$ \;\;\;\; \;\;\;\;\;\; \; J_{c_p}(\et_{q+p})  J_{d_p}(\o_{q+p})
R_g^{\a_1} (\T^1,z_1) \ldots  R_g^{\a_n} (\T^n,z_n) \rangle
 \eqno(3.7) $$
since the required averages are in the affine-Sugawara theory on $g$. The
first-order  connections are
$$  W_{i,0} = 2 \tL^{ab} \sum_{j\neq i}^n  { \T_a^i \T_b^j \over z_{ij} }
\;\;\;\;,\;\;\;\;\; W_{0,i} = 2 L^{ab} \sum_{j\neq i}^n
{\T_a^i \T_b^j \over z_{ij} } \eqno(3.8a) $$
 $$ W_{i,0} + W_{0,i} =W_i^g = 2 L^{ab}_g \sum_{j\neq i}^n
{\T_a^i \T_b^j \over z_{ij} } \eqno(3.8b) $$
where $W_i^g$ are the affine-Sugawara connections obtained by Knizhnik and
Zamo- \linebreak
lodchikov. The second-order connections are given for completeness in
Appendix A.

\noindent 4. Invariant correlators. Under $T$ and $\tT$, the biconformal
correlators enjoy an $SL(2,\R)  \times SL(2,\R)$ covariance, and the invariant
four-point correlators $Y$ are
$$ Y^{\a} (\bu,u) = \left( \prod_{i<j}^4 \bz_{ij}^{\bc_{ij}} z_{ij}^{\g_{ij}}
\right) A^{\a} ( \bz,z)   \;\;\;\;,\;\;\;\;
 u = { z_{12} z_{34} \over z_{14} z_{32} } \;\;\;, \;\;\;\;
\bu = { \bz_{12} \bz_{34} \over \bz_{14} \bz_{32} } \eqno(3.9a)$$
$$ \eqalign{ \g_{12} = \g_{13}  = 0\;\;\;,\;\;\;\; \g_{14} = 2 \,
\D_{\a_1} \;\;\;&,\;\;\;\;
 \g_{23} = \D_{\a_1} + \D_{\a_2} + \D_{\a_3} - \D_{\a_4} \cr
 \g_{24} = -\D_{\a_1} + \D_{\a_2} - \D_{\a_3} + \D_{\a_4} \;\;\; &,\;\;\;\;
 \g_{34} = -\D_{\a_1} - \D_{\a_2} + \D_{\a_3} + \D_{\a_4} \;\;\;\;\;
\cr} \eqno(3.9b)$$
$$ \bc_{ij} = \g_{ij} \ve_{\D \ra \tD}  \eqno(3.9c) $$
where $\a = (\a_1 \a_2 \a_3 \a_4)$. The invariant correlators satisfy the
invariant affine-Virasoro Ward identities
$$ \bpa^q \pa^p Y^{\a} (\bu,u)\ve_{\bu =u}=Y_g^{\b}(u){W_{qp}(u)_{\b}}^{\a}
 \eqno(3.10a) $$
$$ Y_g^{\a} (u) = Y^{\a} (u,u)  \;\;\;\;,\;\;\;\; Y_g^{\b}
( \sum_{i=1}^4 \T_a^i)_{\b}{}^{\a} =0 \;\;\;\;,\;\;a \in g \eqno(3.10b) $$
where $Y_g(u)$  is the invariant affine-Sugawara correlator and $W_{qp}$ are
the invariant affine-Virasoro connections. The first-order invariant
connections are
$$ W_{10} = 2 \tL^{ab} \left( {\T_a^1 \T_b^2 \over u} +
{ \T_a^1 \T_b^3 \over u-1}  \right) \;\;\;\; , \;\;\;\;
 W_{01} = 2 L^{ab} \left( {\T_a^1 \T_b^2 \over u} +
{ \T_a^1 \T_b^3 \over u-1}  \right) \eqno(3.11a) $$
$$ W_{10} + W_{01} = W^g =2 L_g^{ab} \left( {\T_a^1 \T_b^2 \over u} +
{ \T_a^1 \T_b^3 \over u-1}  \right) \eqno(3.11b) $$
and the second-order invariant connections are given in Appendix A.

\noindent 5. Consistency relations. The affine-Virasoro connections satisfy
the consistency relations
$$ (\pa_i + W_i^g) W_{j_1 \ldots j_q,i_1 \ldots i_p} =
W_{j_1 \ldots j_q i ,i_1 \ldots i_p} + W_{j_1 \ldots j_q,i_1 \ldots i_p i }
\eqno(3.12a) $$
$$ (\pa + W^g) W_{qp} = W_{q+1,p} + W_{q,p+1} \eqno(3.12b) $$
where $W_{00} = 1$. The consistency relations were originally derived [14]
from simple properties of the biprimary fields, and the relations are necessary
conditions for factorization, discussed below. In fact, the relations also
follow directly from the definitions (3.5a) and (3.10) of the connections as
derivatives of the biconformal correlators, so the consistency relations are
integrability conditions for the existence of the biconformal correlators. To
understand this, the reader should begin with the slightly simpler case
$f_{qp} \equiv \bpa^q \pa^p f(\bu,u) \ve_{\bu =u} $, which satisfies
$\pa f_{qp} = f_{q+1,p} + f_{q,p+1} $ for all $f(\bu,u)$.

\noindent 6. Factorization. In order to  separate the conformal field theories
of $\tL$ and $L$, we assume the abstract factorization
$$ A^{\a} (\bz,z) = (\bA (\bz)\, A(z) )^{\a} \;\;\;\;,\;\;\;\;
Y^{\a} (\bu,u) = (\bY (\bu)\, Y(u) )^{\a} \eqno(3.13) $$
where the barred and unbarred amplitudes are the proper correlators of the
$\tL$ and the $L$ theories respectively. Then the factorized affine-Virasoro
Ward identities
$$   (\pa_{j_1} \ldots \pa_{j_q} \bA \, \pa_{i_1} \ldots \pa_{i_p} A)^{\a}
= A^{\b}_g  {( W_{j_1 \ldots j_q, i_1 \ldots i_p } )_{\b}}^{\a}
\;\;\;\;,\;\;\;\; A_g^{\a} = ( \bA \, A )^{\a}   \eqno(3.14a)$$
$$ (\pa^q \bY \, \pa^p Y)^{\a}  = Y_g^{\b} {(W_{qp})_{\b}}^{\a}
\;\;\;\;,\;\;\;\; Y_g^{\a} = ( \bY \, Y )^{\a}  \eqno(3.14b) $$
are coupled non-linear differential equations for the K-conjugate pair of
conformal field theories.

To make the factorization concrete, we must also specify the factorized
assignment of the Lie algebra indices. In this paper, we will discuss, at
various levels of completeness, four concrete factorization ans\"atze
$$ \eqalignno{
A^{\a} (\bz,z) & = \sum_{\n} \bA_{\n}^{\b} (\bz) {A_{\n}(z)_{\b}}^{\a}
\;\;\;\;\;\; [\mbox{matrix}] &(3.15a) \cr
 A^{\a} (\bz,z) & = \sum_{\n} \bA_{\n} (\bz) A_{\n}^{\a}(z)
\;\;\;\;\;\;\;\;\;\, [\mbox{vector}] &(3.15b) \cr
 A^{\a} (\bz,z) & = \sum_{\n} \bA_{\n}^{\a} (\bz) A_{\n}(z)
\;\;\;\;\;\;\;\;\;\, [\mbox{vector-bar}] &(3.15c) \cr
 A^{\a} (\bz,z) & = \sum_{\n} \bA_{\n}^{\a} (\bz) A_{\n}^{\a}(z)
\;\;\;\;\;\; \;\;\; [\mbox{symmetric}] &(3.15d) \cr} $$
and the corresponding forms for the invariant amplitudes. The first and last
of these ans\"atze were introduced in [14], where a solution for the coset
constructions was found in the matrix ansatz.  A common feature of these
ans\"atze is the conformal structure index $\n$, which labels the conformal
structures $A_{\n}$. We shall see that a  generically-infinite number of
conformal structures is required to factorize the general affine-Virasoro
construction.

\noindent 7. Coset correlators. The biconformal correlators for $h$ and the
$g/h$ coset constructions are [14]
$$ \tL = L_{g/h} \;\; \; \;  ,\;\; \;\; L = L_h \eqno(3.16a)  $$
$$ A^{\a} (\bz,z)  = A_{g/h}^{\b} (\bz,z_0) {A_h(z,z_0)_{\b}}^{\a}
\eqno(3.16b) $$
$$A_{g/h}^{\a} = A_g^{\b} {(A_h^{-1})_{\b}}^{\a}  \;\;\;\;,\;\;\;\;
 A_{g/h}^{\b} ( \sum_{i=1}^n \T_a^i)_{\b}{}^{\a} =0 \;\;\;\;,\;\;a \in h
 \eqno(3.16c) $$
$$ Y^{\a} (\bu,u)  =Y_{g/h}^{\b}(\bu,u_0) {Y_h(u,u_0)_{\b}}^{\a}\eqno(3.16d)$$
$$ Y_{g/h}^{\a} = Y_g^{\b} {(Y_h^{-1})_{\b}}^{\a} \;\;\;\;,\;\;\;\;
Y_{g/h}^{\b} ( \sum_{i=1}^4 \T_a^i)_{\b}{}^{\a} =0 \;\;\;\;,\;\;a \in h
 \eqno(3.16e)$$
where $A_{g/h}$ and $Y_{g/h}$ are the coset correlators and the two-index
symbols are the invertible evolution operators of $h$,
$$ \pa_i A_h(z,z_0) = A_h (z,z_0) W_i^h (z)\;\;\;\;,\;\;\;\;
 \pa_i A_h^{-1} (z,z_0) = - W_i^h(z) A_h^{-1}  (z,z_0)  \eqno(3.17a) $$
$$  A_h(z_0,z_0{)_{\a}}^{\b} = {1 \over \prod_{i<j}
(z_{ij}^0)^{\g_{ij}^h(\a)} } \d_{\a}^{\b}    \eqno(3.17b) $$
$$ \pa Y_h(u,u_0) =Y_h  (u,u_0) W^h(u) \;\;\;\;,\;\;\;\;
 \pa Y_h^{-1} (u,u_0) = - W^h(u) Y_h^{-1}  (u,u_0)  \eqno(3.17c) $$
$$  Y_h(u_0,u_0{)_{\a}}^{\b} = \d_{\a}^{\b} \;\;\;\;.  \eqno(3.17d) $$
The solution (3.16) resides in the matrix ansatz (3.15a), with only one
conformal structure, and, at the level of conformal blocks, this solution shows
the form
$$ Y_{g/h}^{\a}(u,u_0) = Y_{g/h}^M(u,u_0) v_M^{\a}(h) \;\;\;\;,\;\;\;\;
Y_{g/h}^M(u,u_0) = d^r \C(u)_r{}^R \F_h(u_0)_{R}{}^M  \eqno(3.18a) $$
$$ \C (u)_r{}^R = \F_g(u)_r{}^m \F_h^{-1}(u)_m{}^R \eqno(3.18b) $$
where $v_M^{\a}(h)$ are the $h$-invariant tensors of
$\T^1 \oti \cdots \oti \T^4$. The $u$-dependent factors $\C (u)_r{}^R$ are the
coset blocks defined by Douglas [15,14,19].

An important subtlety here is that the evolution operators $A_h$ and $Y_h$ are
not the $h$ correlators, because they do not satisfy the $h$-global Ward
identities. The proper factorization of (3.16) into the {\it correlators}
of $g/h$ and $h$ is [14]
$$ A^{\a}(\bz,z) = A_{g/h}^M (\bz,z_0) \, A_h(z,z_0)_M{}^{\a} \eqno(3.19a) $$
$$ A_{g/h}^{\a}= A_{g/h}^M \, w_M^{\a}(h) \;\;\;\;,\;\;\;\;
(A_h)_M{}^{\a} \equiv w_M^{\b} (h) (A_h)_{\b}{}^{\a} \eqno(3.19b) $$
$$ Y^{\a}(\bu,u) = Y_{g/h}^M (\bu,u_0) \, Y_h(u,u_0)_M{}^{\a} \eqno(3.19c) $$
$$ Y_{g/h}^{\a}= Y_{g/h}^M \, v_M^{\a}(h) \;\;\;\;,\;\;\;\;
(Y_h)_M{}^{\a} \equiv v_M^{\b} (h) (Y_h)_{\b}{}^{\a} \eqno(3.19d) $$
where $w_M^{\a}(h)$ are the $h$-invariant tensors of
$\T^1 \oti \cdots \oti \T^n$. In (3.19), the projected factors $(A_h)_M{}^{\a}$
and $(Y_h)_M{}^{\a}$ may be identified as $h$ correlators because they
satisfy the $h$-global Ward identities. Moreover, the factors $A_{g/h}^{M}$
and $Y_{g/h}^M$ (see (3.18)) are equivalent representations of the coset
correlators $A_{g/h}^{\a}$ and $Y_{g/h}^{\a}$, since the two sets are equal up
to constant tensors. We finally note that the factorization (3.19) is in the
vector ansatz (3.15b) with $\n=M$.

For use below, we also give
the known connections for $h$ and $g/h$
$$ W_{j_1 \ldots j_q,i_q \ldots i_p} [\tL = L_{g/h} , L= L_h]
= W_{j_1 \ldots j_q,0}^{g/h}   W_{0, i_1 \ldots i_p}^{h}   \eqno(3.20a)  $$
$$  W_{j_1 \ldots j_q j_{q+1},0 }^{g/h}= (\pa_{j_{q+1}} + W^g_{j_{q+1}} )
 W_{j_1 \ldots j_q,0}^{g/h} - W_{j_1 \ldots j_q, 0 }^{g/h} W^h_{j_{q+1}}
 \eqno(3.20b) $$
$$ W_{0, i_1 \ldots i_p i_{p+1} }^h = (\pa_{i_{p+1}} + W^h_{i_{p+1}} )
W_{0, i_1 \ldots i_p }^h  \eqno(3.20c) $$
$$  W_{qp}[\tL = L_{g/h} , L= L_h ] = W_{q0}^{g/h} W_{0p}^h \eqno(3.20d) $$
$$ W_{q+1,0}^{g/h}  = ( \pa + W^g) W_{q0}^{g/h}-W_{q0}^{g/h}W^h \eqno(3.20e) $$
$$ W_{0,p+1}^{h}  = (\pa + W^h )  W_{0p}^{h}   \eqno(3.20f) $$
where $W_{00}= 1$ and the first-order connections of $g$ and $h$ are given in
(3.8), (3.11). These results are rederived in Section 5 by a method which also
generates the connections and conformal correlators of all the affine-Sugawara
nests.

\section{Some General Properties of the A-V Connections}
In this section, we discuss a number of general properties of the
affine-Virasoro (A-V) connections.

\noindent A. Solution of the consistency relations. It was noted in Ref.[14]
that the consistency relations (3.12) can be solved to obtain the general
connections in terms of the canonical sets $W_{0p}$ or $W_{0,i_1 \ldots i_p}$,
or similar sets. The explicit forms of these solutions\footnote{In (4.1a,b),
$\sum_{\P}$ sums over all permutations of the indicated
indices and the following rules are included:
$\prod_{k=1}^0 \cD_{j_k}^g \equiv 1$ and
$W_{0,j_{q+1} \ldots j_q i_1 \ldots i_p} \equiv W_{0,i_1 \ldots i_p} $
for the $r=0,q$ terms in (4.1a), and similarly for the $r=0,p$ terms in
(4.1b).}
$$\eqalignno{  W_{j_1 \ldots j_q,i_1 \ldots i_p}
& = \sum_{r=0}^q (-1)^{q-r} \ch{q}{r} {1 \over q !}
\sum_{\P(j_1 \ldots j_q)} \left[ ( \prod_{k=1}^r \cD_{j_k}^g )
W_{0,j_{r+1} \ldots j_q i_1 \ldots i_p } \right] \;\;\;\; \; &(4.1a) \cr
& = \sum_{r=0}^p (-1)^{p-r} \ch{p}{r} {1 \over p !}
\sum_{\P(i_1 \ldots i_p)} \left[ ( \prod_{k=1}^r \cD_{i_k}^g )
W_{i_{r+1} \ldots i_p j_1 \ldots j_q,0 } \right] &(4.1b) \cr} $$
$$ \cD_{i}^g \equiv \pa_{i} + W_{i}^g $$
$$ \eqalignno{
 W_{qp} & = \sum_{r=0}^q (-)^{r}  \ch{q}{r} (\pa + W^g)^{q-r} W_{0,p+r}
 &(4.1c) \cr &= \sum_{r=0}^p (-)^{r}  \ch{p}{r} (\pa + W^g)^{p-r} W_{q+r,0}
&(4.1d) \cr } $$
are easily checked with the binomial identity
 $ ( {q +1 \atop r}) = ( { q \atop r}) + ( {q \atop r-1} ) $.

In what follows, we refer to $W_{0,i_1 \ldots i_p}$,
$W_{j_1 \ldots j_q,0}$, $W_{0p }$ and $W_{q 0}$ as the {\it one-sided}
connections, and the rest of the connections (e.g. $W_{qp},\, q,p \geq 1$)
as the {\it mixed} connections. It is clear from their definition in (3.7)
that the one-sided connections are functions only of $\tL$ or $L$
$$ W_{j_1 \ldots j_q ,0} (\tL) \;\;\;\;,\;\;\;\;
W_{0,i_1 \ldots i_p} (L) \;\;\;\;,\;\;\;\;  W_{q0}(\tL)\;\;\;\;,\;\;\;\;
W_{0p}(L) \eqno(4.2) $$
and so are associated directly to the $\tL$ or the $L$ theory. This property is
not shared by the mixed connections
 $ W_{j_1 \ldots j_q,i_1 \ldots i_p} (\tL,L) $ and $W_{qp}(\tL,L)$, which
are functions of $\tL$ and $L$.

\noindent B. Connection sum rules. The translation sum rules
$$ \eqalign{  \sum_{r,s=0}^{\infty} { 1 \over r ! \, s !}
\sum_{l_1 \ldots l_r}^n
\sum_{k_1 \ldots k_s}^n  \left[ \prod_{\m=1}^r (z_{l_{\m}} -z^0_{l_{\m}})
\right] & \left[\, \prod_{\n=1}^s (z_{k_{\n}} -z^0_{k_{\n}} )  \right]
W_{l_1 \ldots l_r j_1 \ldots j_q, k_{1} \ldots k_s i_1 \ldots i_p} (z_0) \cr
& = A_g(z,z_0) W_{j_1 \ldots j_q ,i_1 \ldots i_p} (z) \cr} \eqno(4.3a)$$
$$ \sum_{r,s=0}^{\infty} { (u- u_0)^{r+s}  \over r! \, s ! }
W_{r+q,s+p}(u_0) = Y_g(u,u_0) W_{q p} (u)  \eqno(4.3b)$$
relate the connections at different points, where
$$ \pa_i {A_g(z,z_0)_{\a}}^{\b} = {A_g (z,z_0)_{\a}}^{\g}\,
{W_i^g(z)_{\g}}^{\b}
 \;\;\;\;,\;\;\;\; {A_g(z_0,z_0)_{\a}}^{\b} =\d_{\a}^{\b} \eqno(4.4a) $$
$$ \pa_u {Y_g(u,u_0)_{\a}}^{\b} = {Y_g (u,u_0)_{\a}}^{\g} \,
{W^g(u)_{\g}}^{\b}
 \;\;\;\;,\;\;\;\;  {Y_g(u_0,u_0)_{\a}}^{\b} = \d_{\a}^{\b}  \eqno(4.4b)$$
are the invertible evolution operators of the affine-Sugawara construction on
$g$. These identities are obtained by repeated application of the consistency
relations.

\noindent C. K-conjugation covariance. It is clear on inspection of (3.7) that
the connections enjoy the K-conjugation covariance
$$ W_{j_1 \ldots j_q, i_1 \ldots i_p}( \tL,L)
= W_{i_1 \ldots i_p, j_1 \ldots j_q}( L,\tL)  \eqno(4.5a) $$
$$ W_{q p} (\tL,L) = W_{p q} (L,\tL) \eqno(4.5b) $$
under the exchange $\tL \lra L $ of the $\tL$ and $L$ theories. The simpler
covariance of the  one-sided connections
$$
 W_{j_1 \ldots j_q,0} (\tL=L_*) = W_{0, j_1 \ldots j_q} (L=L_*)
\;\;\;\;,\;\;\;\;
 W_{0,i_1 \ldots i_p} (L=L_*) = W_{i_1 \ldots i_p,0} (\tL=L_*) \eqno(4.6a) $$
$$
 W_{q 0}(\tL=L_*) = W_{0 q} (L=L_*)  \;\;\;\;,\;\;\;\;
W_{0 p}(L=L_*) = W_{p 0} (\tL=L_*)   \eqno(4.6b)  $$
follows immediately with (4.2), where $L_*$ is any particular affine-Virasoro
construction.

\noindent D. Crossing symmetry. The computations in (3.7) can be performed
for a fixed ordering of operators and then again after an exchange $ k \lra l$
(including $T$'s, $z$'s and indices). The crossing symmetry of the connections
$$  W_{j_1 \ldots j_q,i_1 \ldots i_p}  \ve_{k \lra l}
=  W_{j_1 \ldots j_q,i_1 \ldots i_p} \eqno(4.7) $$
then follows from the usual analyticity in $z$ for (derivatives of) the
biconformal correlators at $z_i=\bz_i$. We have checked this symmetry for the
explicit forms in (3.8), (A.1) and the affine-Sugawara nest connections
of Section 5 and Appendix B. A corresponding crossing relation of the invariant
connections
$$  W_{q p} (1-u) = (-)^{q+p}  P_{23} W_{q p} (u) P_{23} \eqno(4.8a) $$
$$ P_{23} \T^2 P_{23} = \T^3 \;\;\;\;,\;\;\;\; P_{23}^2 =1 \eqno(4.8b) $$
follows by $SL(2,\R) \times SL(2,\R)$ covariance from (4.7), using (A.3) and
the explicit form of $W_{01}$ in (3.11a).

\noindent E. High-level connections\footnote{ The following discussion assumes
a fixed choice of external representations $\T$ in the biconformal
correlators.}. To leading non-trivial order in $k^{-1}$ on simple $g$, the
currents are effectively abelian and we can evaluate (3.7) by Wick expansion
using [12,17]
$$ L^{ab} = {P^{ab} \over 2k} + \cO (k^{-2}) \;\;\;\;,\;\;\;\;
\tL^{ab} = {\tilde{P}^{ab} \over 2k} + \cO (k^{-2}) \eqno(4.9a) $$
$$ P^{ab} + \tilde{P}^{ab} = \eta^{ab}  \;\;\;\;,\;\;\;\;
P^{ab} \et_{bc} \tilde{P}^{cd} =0 \eqno(4.9b)$$
$$ \undercontract{J_a (z) J_b(w)} = { k \et_{ab} \over (z-w)^2 } + \cO(k^{0})
\;\;\;\; . \eqno(4.9c) $$
Here, $\eta^{ab}$ is the Killing metric on $g$ and
$\tilde{P}$ and $P$ are the high-level projectors of the $\tL$ and $L$
theories respectively. The leading term in (3.7) is obtained at $\cO (k^0)$
from the maximum number $p+q$ of contractions. This contribution vanishes
because no $\o_i - z_j$ linkage is generated (between the currents and the
affine primary fields), so the contour integrals are zero. The next to leading
term at $\cO (k^{-1})$ can be computed with  $p+q-1$ contractions. As seen in
the leading term, the only contractions which survive the contour integrations
are those for which no subset of contractions covers a complete subset of
$(\o_i,\et_i)$ pairs. Then it is not difficult to see that all contributing
terms involve a coefficient of the form
$$ (P_1 \eta P_2 \et \ldots \et P_{p+q})^{ab} \;\;, $$
where $P_r$ may be $P$ or $\tilde{P}$. According to eq.(4.9b), this factor is
zero unless all the projectors are the same, which says that the mixed
connections are $\cO (k^{-2})$.

The high-level form of the affine-Virasoro connections
$$ W_{j_1 \ldots j_q,0} = \left( \prod_{r=1}^{q-1} \pa_{j_r} \right) W_{j_q,0}
+
\cO (k^{-2}) \;\;\;,\;\; q \geq 1 \eqno(4.10a)$$
$$ W_{0,i_1 \ldots i_p} = \left( \prod_{r=1}^{p-1} \pa_{i_r} \right) W_{0,i_p}
+
\cO (k^{-2}) \;\;\;,\;\; p \geq 1 \eqno(4.10b)$$
$$ W_{j_1 \ldots j_q,i_1 \ldots i_p} =\cO (k^{-2})\;\;\;,\;\; q,p \geq 1
 \eqno(4.10c) $$
is then obtained by solving the consistency relations for the one-sided
connections. Similarly, we obtain
$$ W_{q 0} = \pa^{q-1} W_{1 0} + \cO (k^{-2}) \;\;\;,\;\; q \geq
1\eqno(4.11a)$$
$$ W_{0 p} = \pa^{p-1} W_{0 1} + \cO (k^{-2}) \;\;\;,\;\; p \geq
1\eqno(4.11b)$$
$$ W_{q p} = \cO (k^{-2})   \;\;\;, \; \; q,p \geq 1 \eqno(4.11c)$$
for the invariant connections. It follows from (4.10) and (4.11) that both sets
may be written in the factorized forms
$$ W_{j_1 \ldots j_q,i_1 \ldots i_p} = W_{j_1 \ldots j_q,0}
W_{0,i_1 \ldots i_p} + \cO (k^{-2})  \eqno(4.12a)$$
$$ W_{q p} =W_{q 0} W_{0 p} + \cO (k^{-2}) \eqno(4.12b) $$
which, according to eq.(3.20), are exact to all orders for  $\tL = L_{g/h}$ and
$L=L_h$.

\section{The Affine-Sugawara Nests}
In this section, we formulate an iterative procedure, using  K-conjugation
and the solutions (4.1) of the consistency relations, to obtain the
connections and conformal correlators of the affine-Sugawara nests [9,10,11]
$$ \tL = L_{g/h_1 /\ldots / h_n} \;\;\;\;,\;\;\;\;
L = L_{h_1 / \ldots / h_n} \;\;\;\;. \eqno(5.1) $$
For simplicity, we will follow the argument for the invariant correlators and
give the corresponding results for the $n$-point correlators at the end.

In this development, we will use two identities repeatedly
$$ (\pa +W)^p A = Y^{-1} \pa^p (Y A) \;\;\;\;,\;\;\;\; \pa Y(u,u_0) =
Y(u,u_0) W (u) \;\;\;\;,\;\;\;\; Y(u_0,u_0) = \one     \eqno(5.2a) $$
$$ \sum_{r=0}^q (-)^r \ch{q}{r} \pa^{q-r} [ A (\pa^r B) C] =
\sum_{r=0}^q \ch{q}{r}( \pa^{q-r} A)  B (\pa^r C)  \eqno(5.2b) $$
which hold for all  $A,\, B,\,C$ and $ W$, where $Y$ is the invertible
evolution operator of $W$.

We begin with the one-sided connections of the trivial theory
$$ W_{0 p} (L=0) = \one \, \d_{0,p} \;\;\;\; . \eqno(5.3) $$
Substitution of these into the solution (4.1c) gives
$$ W_{q p}  (\tL =L_g,L=0) = (\pa + W^g)^q \one \, \d_{0,p} \eqno(5.4) $$
which implies the one-sided affine-Sugawara connections on $g$
$$ W_{q 0} (\tL = L_g) = (\pa + W^g)^q \one =Y_g^{-1} \pa^q Y_g
\;\;\;\;,\;\;\;\; \pa Y_g = Y_g W^g \;\;,\;\; Y_g(u_0,u_0) = \one
\;\;.\eqno(5.5) $$
Here $W_{10}(L_g) = W^g$ is the invariant affine-Sugawara connection in
(3.11b). Renaming the group, we have for $h \subset g$
$$ W_{q 0} (\tL = L_h) = (\pa + W^h)^q \one = Y_h^{-1} \pa^q Y_h
\;\;\;\;,\;\;\;\; \pa Y_h = Y_h W^h \;\;,\;\; Y_h(u_0,u_0) = \one
\eqno(5.6a) $$
$$ W_{0 p} (L = L_h) = (\pa + W^h)^p \one = Y_h^{-1} \pa^p Y_h  \eqno(5.6b) $$
where (5.6b) follows from (5.6a) and (4.6).
Then, using (5.6b) in the solution (4.1c), we obtain  the connections for
$g/h$ on the left
$$ \eqalign{ \;\;\;\; W_{q p}[\tL = L_{g/h}, L=L_h] & =
\sum_{r=0}^q (-)^r \ch{q}{r} (\pa +W^g)^{q-r} W_{0,p+r}(L=L_h)  \cr
& = \sum_{r=0}^q (-)^r \ch{q}{r} (\pa+ W^g)^{q-r} (\pa +W^h)^{p+r} \one \cr
&= Y_g^{-1} \sum_{r=0}^q (-1)^r \ch{q}{r} \pa^{q-r} (Y_{g/h} \pa^{p+r} Y_h) \cr
& = Y_g^{-1} (\pa^q Y_{g/h} )( \pa^p Y_{h} ) \;\;\;\;,\;\;\;\; \;\;
Y_{g/h} = Y_g Y_h^{-1}  \cr} \eqno(5.7) $$
where the identities (5.2a,b) were used in the last two steps.

At this point in the iteration, we have regained the known results for the
coset constructions, since the one-sided coset connections
$$ W_{q 0} (\tL = L_{g/h} ) =Y_g^{-1} (\pa^q Y_{g/h} ) Y_h \eqno(5.8) $$
solve the recursion relation (3.20e). To see the factorized biconformal
correlators, note that (5.7) may be rewritten as
$$ Y_g^{\b} {W_{qp}[\tL = L_{g/h},L=L_h]_{\b}}^{\a}  =
 \pa^q Y_{g/h}^{\b} \pa^p {(Y_h)_{\b}}^{\a} =
\bpa^q \pa^p Y^{\a} (\bu,u) \ve_{\bu =u}  \eqno(5.9a)$$
$$ Y^{\a} (\bu,u)[\tL = L_{g/h}, L = L_h]
 = Y_{g/h}^{\b} (\bu,u_0)  Y_h(u,u_0)_{\b}{}^{\a} \eqno(5.9b)  $$
$$ Y_{g/h}^{\a} (u,u_0) =Y_g^{\b}(u_0) {Y_{g/h}(u,u_0)_{\b}}^{\a}
\;\;\;\;,\;\;\;\;
Y_g^{\a}(u) = Y_g^{\b}(u_0) {Y_g(u,u_0)_{\b}}^{\a} \;\;. \eqno(5.9c) $$
These forms verify the factorized Ward identities for $g/h$ and $h$, where
$Y_{g/h}^{\a}$ in (5.9c) is the coset correlator. Note also that $Y_{g}^{\a}$,
$W_{qp}$ and the biconformal correlators are independent of the reference point
$u_0$ because
$$ \pa_{u_0} Y(u,u_0) = - W(u_0) Y(u,u_0) \eqno(5.10) $$
follows for the evolution operator in (5.2a).

We move on now to the first nontrivial affine-Sugawara nest, defined on
$g \supset h_1 \supset h_2$. Renaming groups again and using (4.6), we know
from
the result (5.8) for the cosets that
$$ W_{0 p} (L =L_{h_1/h_2})
 =  Y_{h_1}^{-1} (\pa^p Y_{h_1/h_2} ) Y_{h_2}  \;\;\;\;. \eqno(5.11) $$
Then, we obtain the nest connections
$$ \eqalign{ W_{q p} [\tL = L_{g/h_1/h_2} , L=L_{h_1/h_2} ]
& =\sum_{r =0}^q (-)^{r} \ch{q}{r}(\pa+ W^g)^{q-r} W_{0,p+r} (L
=L_{h_1/h_2})\cr
& = \sum_{r=0}^q  (-1)^r \ch{q}{r} (\pa +W^g)^{q-r}[ Y_{h_1}^{-1} ( \pa^{p+r}
Y_{h_1/h_2}) Y_{h_2} ] \cr
& = Y_g^{-1} \sum_{r=0}^q (-1)^r \ch{q}{r} \pa^{q-r}
[Y_{g/h_1} ( \pa^{p+r} Y_{h_1/h_2}) Y_{h_2} ] \cr
& = Y_g^{-1} \sum_{r=0}^q \ch{q}{r}( \pa^{q-r} Y_{g/h_1})( \pa^p Y_{h_1/h_2} )
(\pa^r Y_{h_2} ) \cr} \eqno(5.12) $$
from (4.1c),  where the identities (5.2a,b) were used in the last two steps.
The nest connections  may also be expressed in the form
$$ W_{qp} [\tL = L_{g/h_1/h_2} , L=L_{h_1/h_2} ]
= \sum_{r=0}^q \ch{q}{r} W_{0,q-r}^{g/h_1} W_{0p}^{h_1/h_2} W_{0r}^{h_2}
\eqno(5.13) $$
where $W_{0p}^{g/h} = W_{0p}(L_{g/h})$ and $W_{0p}^h = W_{0p} (L_h)$ are
the one-sided connections of $g/h$ and $h$. In this form, we see that
the nest connections are independent of the reference point $u_0$.

We may also display the nest connections in the form
$$ Y_g W_{q p} [\tL= L_{g/h_1/h_2},L=L_{ h_1/h_2 }]
  = \left\{ \bpa^q \pa^p [ Y_{g/h_1} (\bu) Y_{h_1/h_2} (u)
Y_{h_2} (\bu) ] \right\} \ve_{\bu =u}  \eqno(5.14)$$
which verifies the consistency relations (3.12b) on inspection, according to
the remarks below (3.12). Moreover, the form (5.14) shows that the unfactorized
Ward identities (3.10) are solved by the biconformal nest correlators
$$ Y^{\a}(\bu,u) [\tL = L_{g/h_1/h_2} , L=L_{h_1/h_2} ]
 = Y_{g/h_1}^{\b}(\bu,u_0) \, Y_{h_1/h_2}(u,u_0{)_{\b}}^{\g}
\, Y_{h_2}(\bu,u_0{)_{\g}}^{\a} \eqno(5.15) $$
where $Y_{g/h}^{\a}$ is the invariant coset correlator defined in (3.16e).
According to (5.10), the biconformal nest correlator is independent of the
reference point $u_0$.

In order to factorize the biconformal nest correlator (5.15) into the
conformal correlators\footnote{The evolution operator
$Y_{h_1/h_2}(u,u_0)_{\b}{}^{\g}$ in (5.15) is not a coset correlator because
it does not satisfy the $h_2$-global Ward identities [14].}
of $g/h_1/h_2$ and $h_1/h_2$, we need the expansions [14]
$$ Y_{g/h_1}^{\b} = Y_{g/h_1}^{m_1}  v^{\b}_{m_1}(h_1) \eqno(5.16a) $$
$$ (Y_{h_1/h_2})_{m_1}{}^{\g} \equiv
v^{\b}_{m_1}(h_1) (Y_{h_1/h_2} )_{\b}{}^{\g}=
  (Y_{h_1/h_2})_{m_1}{}^{m_2} v^{\g}_{m_2}(h_2) \eqno(5.16b) $$
$$ (Y_{h_2})_{m_2}{}^{\a} \equiv v^{\g}_{m_g}(h_2) (Y_{h_2})_{\g}{}^{\a}
\eqno(5.16c) $$
$$ (Y_{h_1/h_2})_{m_1}{}^{\b} (\sum_{i=1}^4 \T_a^i)_{\b}{}^{\a} =
 (Y_{h_2})_{m_2}{}^{\b} (\sum_{i=1}^4 \T_a^i)_{\b}{}^{\a} =0
\;\;\;\;,\;\; a \in h_2  \eqno(5.16d) $$
where $v^{\a}_{m_{i}}(h_{i})$ are the $h_{i}$-invariant tensors of
$\T^1 \oti \cdots \oti \T^4$. Then, (5.15) factorizes as follows,
$$ \eqalign{ \;\;\;\;
 Y^{\a}(\bu,u) & [\tL = L_{g/h_1/h_2}   , L=L_{h_1/h_2} ] \cr
 & = \bY(\bu,u_0) [\tL = L_{g/h_1/h_2}]^{m_1}{}_{m_2}{}^{\a} \,
Y(u,u_0)[ L=L_{h_1/h_2} ]_{m_1}{}^{m_2} \;\;\;\;\;\;\;\; \cr} \eqno(5.17a) $$
$$ \eqalign{ \;\;\;\;
 \bY(\bu,u_0) [\tL = L_{g/h_1/h_2}]^{m_1}{}_{m_2}{}^{\a}
  = (Y_{g/h_1} & (\bu,u_0) \otimes Y_{h_2}(\bu,u_0) )^{m_1}{}_{m_2}{}^{\a} \cr
& = Y_{g/h_1}^{m_1}(\bu,u_0) \, Y_{h_2}(\bu,u_0{)_{m_2}}^{\a} \;\;\;\;\;\;\;\;
\cr} \eqno(5.17b) $$
$$  Y(u,u_0)[ L=L_{h_1/h_2} ]_{m_1}{}^{m_2} = Y_{h_1/h_2}(u,u_0{)_{m_1}}^{m_2}
\eqno(5.17c)  $$
where (5.17) solves the factorized Ward identities (3.14b). The factorization
(5.17) is correct for the $h_1/h_2$ theory because the projected factor
$(Y_{h_1/h_2})_{m_1}{}^{\g}=(Y_{h_1/h_2})_{m_1}{}^{m_2}v_{m_2}^{\g}(h_2) $
satisfies the $h_2$-global Ward identities (5.16d). Moreover,
$Y_{g/h_1}^{m_1}$ and $(Y_{h_2})_{m_2}{}^{\a}$ are equivalent representations
of the $g/h_1$ and $h_2$ correlators respectively.

It is then clear from (5.17b) that the conformal field theory of the nest
$L_{g/h_1/h_2}$ is the tensor-product theory $(g/h_1)\otimes h_2$, as
anticipated in Section 2. Appendix C gives the explicit form of the nest
correlators in terms of conformal blocks.

The procedure followed above can be further iterated to obtain the connections
and biconformal correlators of all the affine-Sugawara nests. The general
scheme is
$$ \eqalign{ W_{q0} &(\tL = L_{g/h_1/\ldots /h_n} )  \ra
 W_{0p}(L = L_{h_1/\ldots /h_{n+1}} ) \cr & \ra
 W_{qp}[\tL = L_{g/h_1/\ldots /h_{n+1}}, L = L_{h_1/\ldots /h_{n+1}} ]
  \ra W_{q0} (\tL = L_{g/h_1/\ldots /h_{n+1}})  \cr} \eqno(5.18) $$
where the first step uses (4.6b) and a group relabelling, the second step
uses the solution (4.1c) of the invariant consistency relations, and the last
step uses (4.2) at $p=0$. Continuing the iteration, we find the invariant
biconformal nest correlators
$$  \eqalign{ Y^{\a} & (\bu,u)
[\tL = L_{g/h_1 / \ldots /h_{2n+1}}, L = L_{h_1 / \ldots /h_{2n+1}}]  = \cr
 & Y_{g/h_1}^{\b} (\bu,u_0) [Y_{h_1/h_2} (u,u_0) Y_{h_2/h_3} (\bu,u_0)
  \cdots Y_{h_{2n}/h_{2n+1}}(\bu,u_0)
Y_{h_{2n+1}} (u,u_0) {]_{\b}}^{\a}  \cr}  \eqno(5.19a)$$
$$  \eqalign{ Y^{\a} & (\bu,u)
[\tL = L_{g/h_1 / \ldots /h_{2n}}, L = L_{h_1 / \ldots /h_{2n}}]  = \cr
 & Y_{g/h_1}^{\b} (\bu,u_0) [ Y_{h_1/h_2} (u,u_0)Y_{h_2/h_3} (\bu,u_0)
  \cdots Y_{h_{2n-1}/h_{2n}}(u,u_0)
Y_{h_{2n}} (\bu,u_0) {]_{\b}}^{\a}  \cr}  \eqno(5.19b)$$
which are independent of $u_0$ and solve the unfactorized Ward identities
(3.10).

The same scheme can be followed to obtain the $n$-point biconformal nest
correlators
$$ \eqalign{  \;\;\;\;\; A^{\a} (\bz,z) &
[\tL = L_{g/h_1 / \ldots /h_{2n+1}}, L = L_{h_1 / \ldots /h_{2n+1}}] =   \cr
 A_{g/h_1}^{\b} & (\bz,z_0)
[ A_{h_1/h_2}(z,z_0) A_{h_2/h_3}(\bz,z_0) \cdots A_{h_{2n}/h_{2n+1}}(\bz,z_0)
A_{h_{2n+1}}(z,z_0){]_{\b}}^{\a} \;\;\;\; \cr} \eqno(5.20a)$$
$$ \eqalign{ \;\;\;\;\; A^{\a} (\bz,z) &
[\tL = L_{g/h_1 / \ldots /h_{2n}}, L = L_{h_1 / \ldots /h_{2n}}] =   \cr
 A_{g/h_1}^{\b} & (\bz,z_0)
[ A_{h_1/h_2}(z,z_0)A_{h_2/h_3}(\bz,z_0) \cdots A_{h_{2n-1}/h_{2n}}(z,z_0)
A_{h_{2n}}(\bz,z_0){]_{\b}}^{\a}  \cr} \eqno(5.20b)$$
which are independent of the reference point and solve the unfactorized Ward
identities (3.5). Using the general principles of the iteration, the results
(5.19) and (5.20) are verified in Appendix B.

Following the discussion of the first nest above, Appendix C discusses the
factorization of the biconformal nest correlators (5.19) into the conformal
correlators of the general nest. The result
$$  \eqalignno{
 \bY [\tL = L_{g/h_1 / \ldots /h_{2n+1}}] & = Y_{g/h_1} \otimes Y_{h_2/h_3}
 \otimes  \cdots \otimes Y_{h_{2n-2}/h_{2n-1}}
\otimes Y_{h_{2n}/h_{2n+1}} \;\;\;\;\;\;\;\;\; &(5.21a) \cr
 \bY [\tL = L_{g/h_1 / \ldots /h_{2n}}] & = Y_{g/h_1} \otimes Y_{h_2/h_3}
  \otimes \cdots \otimes
Y_{h_{2n-2}/h_{2n-1}} \otimes Y_{h_{2n}} \;\;\;\;\;\;\;\;\; &(5.21b) \cr} $$
shows that all the affine-Sugawara nests are tensor-product theories. The
conformal blocks of the general nest and an explicit example are also worked
out in Appendix C.

Beyond the coset constructions and affine-Sugawara nests, it is clear that
solution of the factorized Ward identities will be more complex. In the
following sections, we develop an algebraic reformulation of the system, which,
given the affine-Virasoro connections, allows the construction of global
solutions across all affine-Virasoro space.

\section{Algebraization of the Ward Identities}

Given the affine-Virasoro connections, the factorized affine-Virasoro Ward
identities (3.14) are an all-order system of  non-linear differential
equations. In this section we show that the system has an equivalent
algebraic formulation, observed in discussion with E. Kiritsis.

The algebraization may be understood in two ways. In the first viewpoint, we
solve the unfactorized Ward identities (3.5) and (3.10) by the
partially-factorized forms of the biconformal correlators
$$ \eqalign{ A^{\a} & (\bz,z) = \cr
 & \sum_{q,p=0}^{\infty} \frac{1}{q!}
\sum_{j_1 \ldots j_q}^n \frac{1}{p!}
\sum_{i_1 \ldots i_p}^n \prod_{\m=1}^q (\bz_{j_{\m}} - z^0_{j_{\m}}) \,
[ A_g^{\b}(z_0) W_{j_1 \ldots j_q,i_1 \ldots i_p}(z_0 {)_{\b}}^{\a}]
 \prod_{\n=1}^p (z_{i_{\n}} -z^0_{i_{\n}} ) \cr} \eqno(6.1a) $$
$$ Y^{\a}(\bu,u) = \sum_{q,p=0}^{\infty} {(\bu -u_0)^q \over q!}\,
[Y_g^{\b}(u_0) W_{qp} (u_0 {)_{\b}}^{\a} ]\,{(u-u_0)^p \over p!}  \eqno(6.1b)$$
where $z_0 = \{ z_i^0 \}$ and $u_0$ are regular reference points. These forms
verify the unfactorized Ward identities by differentiation, as
seen explicitly for the invariant case as follows
$$ \eqalign{ \bpa^q \pa^p  Y^{\a}(\bu,u) \ve_{\bu= u} & =
\sum_{r,s=0}^{\infty} { (u-u_0)^{r+s} \over r! \, s! } \,Y_g^{\b} (u_0) \,
W_{r+q,s+p} (u_0 {)_{\b}}^{\a} \cr
& = Y_g^{\b}(u_0) ( Y_g(u,u_0) \,  W_{qp} (u) {)_{\b}}^{\a} =
Y_g^{\b} (u) \, W_{qp} (u {)_{\b}}^{\a} \;\;\;\;. \cr } \eqno(6.2) $$
Here, the translation sum rules (4.3b) were used in the last step, and the
same steps with (4.3a) verify the partially-factorized form of the general
biconformal correlators in (6.1a).

Note also that the biconformal correlators in (6.1) are independent of the
reference point used to define the partial factorization, for example,
$$ \eqalign{
\pa_{u_0} Y^{\a} & (\bu,u)  =\sum_{q,p=0}^{\infty} {(\bu -u_0)^q \over q!}
\, {(u-u_0)^p \over p!} \, Y_g^{\b}(u_0) \cr
& \times  [ (\pa_{u_0} +W^g(u_0)) W_{qp} (u_0) -W_{q+1,p} (u_0) -
W_{q,p+1} (u_0) {]_{\b}}^{\a}   =0 \cr}  \eqno(6.3)$$
where the consistency relations (3.12b) were used in the last step.
Similarly, $\pa / \pa z_i^0 A^{\a} (\bz,z) =0$ is verified with the
consistency relations (3.12a).

The meaning of the partially-factorized forms in (6.1) is that the factorized
Ward identities can be solved algebraically. More precisely, the biconformal
correlators are completely factorized (and hence the factorized Ward identities
solved) if we can factorize the connections at the reference point, which is
an algebraic problem. For the general connections, the abstract form of this
factorization reads
$$ C_{j_1 \ldots j_q,i_1 \ldots i_p}^{\a} \equiv
A_g^{\b} (z_0) \, W_{j_1 \ldots j_q,i_1 \ldots i_p}(z_0 {)_{\b}}^{\a} =
(\bA_{j_1 \ldots j_q} \, A_{i_1 \ldots i_p} )^{\a} \eqno(6.4a)$$
$$ \bA_{j_1 \ldots j_q} \equiv \pa_{j_1} \ldots \pa_{j_q} \bA(z) \ve_{z=z_0}
 \;\;\;\;,\;\;\;\; A_{i_1 \ldots i_p} \equiv \pa_{i_1} \ldots \pa_{i_p} A(z)
\ve_{z=z_0} \eqno(6.4b)$$
while for the invariant case we have the simpler problem
$$ C^{\a}_{qp } \equiv  Y^{\b}_g(u_0) \, W_{qp} (u_0 {)_{\b}}^{\a} =
(\bY_q \, Y_p)^{\a} \eqno(6.5a)$$
$$ \bY_q \equiv \pa^q \bY(u) \ve_{u=u_0} \;\;\;\;,\;\;\;\;
Y_p \equiv \pa^p Y(u) \ve_{u=u_0} \;\;\;\;. \eqno(6.5b) $$
In the following section, we shall return  to study the concrete factorization
ans\"atze (3.15) in this algebraic form.

An equivalent statement of the algebraization is as follows. The factorized
Ward identities in (3.14) are completely solved if they are solved at the
reference point, where they read
$$ (\bA_{j_1 \ldots j_q} \,  A_{i_1 \ldots i_p} )^{\a} =
C_{j_1 \ldots j_q,i_1 \ldots i_p}^{\a} \eqno(6.6a) $$
$$ (\bY_q \, Y_p)^{\a}  =  C^{\a}_{qp }  \eqno(6.6b)$$
in the notation of (6.4) and (6.5). To check this for the invariant case,
assume (6.6b) and follow the steps
$$\eqalign{
 (\pa^q \bY \, \pa^p Y )^{\a} &= \sum_{r=0}^{\infty} {(u-u_0)^r \over r!}
\pa^r (\pa^q \bY \, \pa^p Y )^{\a} \ve_{u=u_0} \cr
 & = \sum_{r=0}^{\infty} {(u-u_0)^r \over r!} \sum_{s=0}^r \ch{r}{s}
 (\pa^{q +s}\bY \, \pa^{p+r-s} Y )^{\a} \ve_{u=u_0} \cr
& = \sum_{r=0}^{\infty} {(u-u_0)^r \over r!} \sum_{s=0}^r \ch{r}{s}
 (\bY_{q+s} \, Y_{p+r-s} )^{\a}  \cr
 & = \sum_{r=0}^{\infty} {(u-u_0)^r \over r!} \sum_{s=0}^r \ch{r}{s}
 C_{q +s, p+r-s}^{\a}  \cr
 & = Y_g^{\b} (u_0) \sum_{r=0}^{\infty} {(u-u_0)^r \over r!}
 \sum_{s=0}^r \ch{r}{s} W_{q +s, p+r-s}(u_0{)_{\b}}^{\a}   \cr
 & = Y_g^{\b} (u_0) \sum_{r,s=0}^{\infty} {(u-u_0)^{r+s} \over r! \, s!}
W_{q +s, p+r}(u_0{)_{\b}}^{\a} \cr
 & = Y_g^{\b} (u) W_{qp}(u {)_{\b}}^{\a} \cr}  \eqno(6.7) $$
where
$ \sum_{r=0}^{\infty} \sum_{s=0}^r f(r,s) = \sum_{r,s=0}^{\infty} f(r+s,s)$
and the translation sum rule (4.3b) were used in the final steps.
Similarly, one uses the translation sum rule (4.3a) to see that the factorized
$n$-point Ward identities (3.14a) are solved by the algebraic factorization
(6.6a).

\section{Factorization}

In this section, we factorize the invariant biconformal correlators via
concrete realizations of the algebraic factorization (6.5).

In particular, we distinguish four concrete algebraic factorization ans\"atze
$$ \eqalignno{ W_{qp}(u_0 {)_{\b}}^{\a} & = \sum_{\n} (\bY_{q\n} {)_{\b}}^{\g}
(Y_{\n p} {)_{\g}}^{\a}
\;\;\;\;\;\; [\mbox{matrix}] &(7.1a) \cr
W_{qp}(u_0 {)_{\b}}^{\a} & = \sum_{\n} \bY_{q\n\b} \,   Y_{\n p}^{\a}
\;\;\;\;\;\;\;\;\;\, \;\;\;\; \; \,\,  [\mbox{vector}] &(7.1b) \cr
W_{qp}(u_0 {)_{\b}}^{\a} & = \sum_{\n} \bY_{q\n\b}^{\a} \,   Y_{\n p}
\;\;\;\;\;\;\;\;\;\, \;\;\;\; \; \,\,  [\mbox{vector-bar}] &(7.1c) \cr
 C_{qp}^{\a}  & = \sum_{\n} \bY_{q\n}^{\a} \, Y_{\n p}^{\a}
\;\;\;\;\;\; \;\;\;   \;\;\;\;\; \;\;\; \,
[\mbox{symmetric}] &(7.1d) \cr} $$
which correspond to the factorization ans\"atze listed in (3.15) as follows,
$$ \eqalignno{
[\mbox{matrix}] \;\;\;\;\; \;\;\;\;\; \;\;\;\;\; \;\;
\;\;\;\;\; \;\;\;\;\; \;\;\;\;\,
  Y^{\a} (\bu,u)   = \sum_{\n} &  \bY_{\n}^{\b} (\bu)  Y_{\n}(u {)_{\b}}^{\a}
    &(7.2a)   \cr
\bY_{\n}^{\a} (\bu) = Y_g^{\b} (u_0)  \sum_{q=0}^{\infty}
{( \bu -u_0)^q \over q!} \,  (\bY_{q \n} {)_{\b}}^{\a}   \;\;\;& , \;\;\;
Y_{\n} ( u {)_{\b}}^{\a} = \sum_{p=0}^{\infty}  (Y_{\n p} {)_{\b}}^{\a} \,
{( u -u_0)^p \over p!}  & \cr
 [\mbox{vector}]  \;\;\;\;\; \;\;\;\;\;\; \;\;\;\;\; \;\;
\;\;\;\;\; \;\;\;\;\; \;\;\;\;
Y^{\a} (\bu,u)   = \sum_{\n} & \bY_{\n} (\bu)  Y^{\a}_{\n}(u)  &(7.2b)  \cr
\bY_{\n} (\bu) =Y_g^{\b} (u_0)  \sum_{q=0}^{\infty} {( \bu -u_0)^q \over q!}
 \, \bY_{q \n \b}    \;\;\;&,\;\;\;
Y_{\n}^{\a} ( u ) = \sum_{p=0}^{\infty}  Y_{\n p}^{\a} \,
 {( u -u_0)^p \over p!}  & \cr
[\mbox{vector-bar}]  \;\;\;\;\; \;\;\;\;\;\; \;\;\;\;\; \;\;
\;\;\;\;\; \;\;\;
Y^{\a} (\bu,u)   = \sum_{\n} & \bY_{\n}^{\a} (\bu)  Y_{\n}(u)
   &(7.2c)   \cr
\bY_{\n}^{\a}(\bu) =Y_g^{\b} (u_0)\sum_{q=0}^{\infty} {( \bu -u_0)^q \over q!}
 \, \bY_{q \n \b}^{\a}    \;\;\;&,\;\;\;
Y_{\n}( u ) = \sum_{p=0}^{\infty}  Y_{\n p} \,
 {( u -u_0)^p \over p!} &  \cr
 [\mbox{symmetric}]  \;\;\;\;\; \;\;\;\;\;\; \;\;\;\;\; \;\;
\;\;\;\;\; \;\;\,
Y^{\a} (\bu,u) = \sum_{\n} & \bY_{\n}^{\a} (\bu)  Y_{\n}^{\a} (u ) &(7.2d)  \cr
\bY_{\n}^{\a} (\bu) = \sum_{q=0}^{\infty} {( \bu -u_0)^q \over q!} \,
   Y_{q \n}^{\a}   \;\;\;& ,\;\;\;
Y_{\n}^{\a} ( u ) = \sum_{p=0}^{\infty} Y_{\n p}^{\a} \,
 {( u -u_0)^p \over p!} \;\;\;\;. & \cr} $$
The four ans\"atze share the notion of a conformal structure index $\n$, while
differing in the assignment of the Lie algebra indices $\a,\,\b$. We remind the
reader that the solution (3.19) for $g/h$ and $h$ resides in the vector
ansatz (7.2b) with $\n=M$. Moreover, the factorization (5.17a) of the first
non-trivial affine-Sugawara nest is in the vector-bar ansatz with
$\n=(m_1,m_2)$. For the general nest, the factorization of (5.19) (see
eq.(C.3)) is in the vector ansatz for $\tL=L_{g/h_1/ \ldots/h_{2n+1}}$
and in the vector-bar ansatz for $\tL=L_{g/h_1/ \ldots/h_{2n}}$.

More generally, the matrices $W_{qp} (u_0)  $ and $ C_{qp} = Y_g (u_0)
W_{qp} (u_0) $ are infinite dimensional, so we expect (and will find) that each
of the ans\"atze exhibits infinite-dimensional factorizations, with an infinite
number of conformal structures, for any K-conjugate pair of affine-Virasoro
constructions.
An infinite-dimensional conformal structure is expected in irrational conformal
field theory, but the problem is that there are too many solutions, many of
which are apparently not physical.

As an example, consider the matrix ansatz (7.1a), whose solutions for any
invertible $\bY$ are
$$ Y_{\n p } = \sum_{q=0}^{\infty} ( \bY^{-1})_{ \n q} W_{qp} (u_0)
\;\;\;\;, \;\;\; \n, p = 0,1,\ldots \;\;\;\;. \eqno(7.3) $$
This is a very large class of solutions to the Ward identities, most of which
must be unacceptable as they stand. To understand this, consider the simple
particular solution
$$ (\bY_{ q \n } {)_{\a}}^{\b} =(\bY^{-1})_{q\a ,\n} {}^{\b} =
 \d_{q \n } \d_{\a}^{\b} \;\;\;\;,\;\;\;\;
(Y_{\n p } {)_{\a}}^{\b} = W_{  \n p } (u_0 {)_{\a}}^{\b} \eqno(7.4a) $$
$$ \bY_{\n}^{\a} (u,u_0) =  Y_g^{\a} (u_0) \,
{( u -u_0)^{\n} \over \n !} \;\;\;\;,\;\;\;
Y_{\n} (u,u_0 {)_{\a}}^{\b}  =  \sum_{p =0}^{\infty}
W_{ \n p } (u_0 {)_{\a}}^{\b} \, {( u -u_0)^p \over p!}  \eqno(7.4b) $$
whose conformal structures $\bY_{\n}^{\a},\, \n = 0,1, \ldots $ do not
show the conformal weights of the $\tL$ theory. We believe that these conformal
structures  should be viewed only as a {\it basis} for a physical solution,
reasoning as follows. Given any particular solution $\bY (u_0),\,Y(u_0)$ to
(7.1a), we also obtain the  associated family of solutions
$$ \bY^{\O} (u,u_0) = \bY (u,u_0)\, \O (u_0)  \;\;\;\;, \;\;\;\;
Y^{\O} (u,u_0) = \O^{-1} (u_0) \, Y (u,u_0) \eqno(7.5) $$
where $\bY (u,u_0),\, Y(u,u_0)$ is the particular solution and $\O (u_0) $ is
an arbitrary invertible matrix. It is then clear that the conformal structures
$\bY_{\n}^{\a} (u,u_0)$  in (7.4) are a basis for the family
$$  (\bY_{\m}^{\a})^{\O}(u,u_0) =
\sum_{\n=0}^{\infty}  \bY_{\n}^{\a} (u,u_0) \, \O (u_0)_{\n \m} =
\sum_{\n=0}^{\infty} Y_g^{\a} (u_0)\,
{ (u-u_0)^{\n} \over \n !} \, \O (u_0)_{\n \m} \eqno(7.6) $$
which is an essentially arbitrary power series in   $(u-u_0)$.

Our attention is then focused on the problem of finding a good basis, in which
the  solution is physical, by paying attention to general principles. In what
follows, we study a  natural factorization in the vector ansatz which gives a
global solution across all affine-Virasoro space. This solution
\begin{enumerate}
\item[a)] reproduces the correlators (3.16e) and (5.21) of the coset
constructions and the affine-Sugawara nests,
\item[b)] exhibits braiding for all affine-Virasoro constructions, and
\item[c)] shows physical behavior at high  level for all affine-Virasoro
constructions on simple $g$.
\end{enumerate}

\section{Factorization by Connection Eigenvectors}

The invariant affine-Virasoro connections $ (W_{qp} {)_{\a}}^{\b} $ define
an infinite-dimensional eigenvalue problem
$$  \eqalignno{ \sum_p W_{qp} (u_0{)_{\a}}^{\b} \bps_{p \b}^{(\n)} (u_0) & =
E_{\n} (u_0) \bps_{ q \a}^{(\n)} (u_0) &(8.1a) \cr
\sum_q \ps_{q (\n)}^{ \b }(u_0)  W_{qp} (u_0 {)_{\b}}^{\a}  & =
E_{\n} (u_0) \ps_{p(\n) }^{ \a} (u_0)  &(8.1b) \cr} $$
whose eigenvectors provide a natural factorization in the vector ansatz (7.2b),
$$ W_{qp} (u_0 {)_{\a}}^{\b} = \sum_{\n =0}^{\infty}  \bps_{q \a}^{(\n)} (u_0)
E_{\n} (u_0) \, \ps_{p (\n)}^{ \b } (u_0) \eqno(8.2a) $$
$$ \bY_{\n} (u,u_0) = \sqrt{E_{\n}(u_0)} \, Y_g^{\a} (u_0)  \bps_{\a}^{(\n)}
(u,u_0)    \;\;\;\;,\;\;\;\; \bps_{\a}^{(\n)} (u,u_0) \equiv\sum_{q=0}^{\infty}
 { (u -u_0)^q \over q!} \,\bps_{q \a}^{(\n)} (u_0) \eqno(8.2b) $$
$$  Y_{\n}^{\a} (u, u_0) = \sqrt{E_{\n}(u_0)} \, \ps_{(\n)}^{\a} (u,u_0)
\;\;\;\;,\;\;\;\;  \ps_{(\n)}^{\a} (u,u_0) \equiv \sum_{p=0}^{\infty}
{ (u-u_0)^p \over p!} \,\ps_{p(\n)}^{ \a } (u_0) \;\;\;.  \eqno(8.2c) $$
More precisely, the spectral resolution (8.2a) holds when
$( W_{qp} {)_{\a}}^{\b} $ is diagonalizable, which we shall see is true at
least down to some finite level because it is true at high level (see Section
10). In what follows, we refer to the basic structures
$\bps_{\a}^{(\n)} (u,u_0),\, \ps_{(\n)}^{\a} (u,u_0) $ in (8.2b,c) as the
{\it conformal eigenvectors} of the $\tL$ and $L$ theories respectively. Note
also that only the conformal eigenvectors with $E_{\n} \neq 0$ contribute to
the factorized correlators $\bY$ and $Y$.

An equivalent form of the factorized correlators (8.2b,c)
$$ \eqalignno{
\bY_{\n} (u,u_0) & ={1 \over \sqrt{E_{\n}(u_0)} } \,  \sum_{p=0}^{\infty}
\bps_{p \a}^{(\n)} (u_0) \,   \pa_{u_0}^p \fb_0^{\a} (u,u_0)    &(8.3a) \cr
 Y_{\n}^{\a}(u,u_0) & = {1 \over \sqrt{E_{\n}(u_0)} } \,
\sum_{q=0}^{\infty} \ps_{q(\n)}^{\b } (u_0)
[ (\pa_{u_0} + W^g (u_0) )^q f_0(u,u_0) {]_{\b}}^{\a} &(8.3b) \cr} $$
$$  \eqalignno{ \fb_0^{\a} (u,u_0) & \equiv Y_g^{\b} (u_0)
 \sum_{q=0}^{\infty} { (u-u_0)^q \over q!}\,
 W_{q 0} (u_0 {)_{\b}}^{\a}  &(8.3c) \cr
 f_0 (u,u_0 {)_{\b}}^{\a}  & \equiv \sum_{p=0}^{\infty} W_{0 p}
(u_0 {)_{\b}}^{\a} \,{ (u-u_0)^p \over p!} &(8.3d) \cr }  $$
is obtained by using the eigenvalue equations for $E_{\n} \neq 0$ and the
identities
$$ \eqalignno{ \sum_{q=0}^{\infty} { (u-u_0)^q \over q!}\,
 W_{q p} (u_0 ) & = (\pa_{u_0} + W^g(u_0))^p
\sum_{q=0}^{\infty} { (u-u_0)^q \over q!} \, W_{q 0} (u_0 ) &(8.4a) \cr
\sum_{p=0}^{\infty}  W_{q p} (u_0 ) \, { (u-u_0)^p \over p!} & =
(\pa_{u_0} + W^g(u_0))^q
\sum_{p=0}^{\infty} W_{0 p} (u_0 )\, { (u-u_0)^p \over p!} &(8.4b)\cr} $$
which follow from the consistency relations (3.12b) or their solutions
in (4.1c,d). In this form, the factorized correlators are expressed as
eigenvector projections of the basic structures $\fb_0, \, f_0$.

As a first test of the global solution (8.2) and (8.3), we reconsider the
familiar case $\tL=L_{g/h}$ and $L=L_h$, for which the basic structures
(8.3c,d) are easily summed,
$$ \fb_0^{\a} (u,u_0) = Y_{g/h}^{\a} (u,u_0)  \;\;\;\;, \;\;\;\;
f_0 (u,u_0 {)_{\a}}^{\b}  = Y_h  (u,u_0{)_{\a}}^{\b}   \eqno(8.5) $$
using the connections in (5.8) and (5.6b). Moreover, we can use (5.10) to
evaluate the $u_0$ derivatives in (8.3a), which gives
$$ \bY_{\n}^{g/h} (u,u_0) = Y_{g/h}^{\a} (u,u_0) d_{\a}^{(\n)} (u_0)
\;\;\;\;, \;\; d_{\a}^{(\n)} (u_0) \equiv {1 \over \sqrt{E_{\n}(u_0)} } \,
 \sum_{p=0}^{\infty}
  W_{0p}^h (u_0 {)_{\a}}^{\b}   \bps_{p \b}^{(\n)} (u_0)  \eqno(8.6)$$
for the coset constructions. In this case, the conformal structures are
{\it degenerate} in that all the $u$ dependence of each structure is in the
same correct coset factor $Y_{g/h}^{\a} (u,u_0) $, defined in (3.16e).

We have also checked that the global solution reproduces the known results in
Section 5 for all the affine-Sugawara nests. As an example, the basic
structure $\fb_0$ and the factorized correlators of the first non-trivial
nests,
$$ \fb_0^{\a} (u,u_0) = Y_{g/h_1}^{\b}(u,u_0) Y_{h_2} (u,u_0 {)_{\b}}^{\a}
\eqno(8.7a)$$
$$ \eqalignno{
\bY_{\n}^{g/h_1/h_2} (u,u_0) & =  Y_{g/h_1}^{m_1} (u,u_0)
Y_{h_2}(u,u_0)_{m_2}{}^{\a} D_{m_1 \a}^{(\n)m_2} (u_0)
 &(8.7b) \cr
D_{m_1 \a}^{(\n)m_2} (u_0) & \equiv {1 \over \sqrt{E_{\n}(u_0)} }
  \sum_{p=0}^{\infty}   W_{0p}^{h_1/h_2} (u_0)_{m_1}{}^{m_2}
 \bps_{p \a}^{(\n)} (u_0)  &(8.7c) \cr} $$
$$ v_{m_1}^{\b} (h_1) (W_{0p}^{h_1/h_2} )_{\b}{}^{\a} \equiv
(W_{0p}^{h_1/h_2} )_{m_1}{}^{m_2} v_{m_2}^{\a} (h_2) \eqno(8.7d) $$
are obtained with (8.3), (5.12), (5.10), (3.20e) and (5.16) for
$\tL=L_{g/h_1/h_2}$. The conformal structures in (8.7b) are again degenerate,
with all $u$ dependence in the correct nest factor $Y_{g/h_1}\oti Y_{h_2}$.

In what follows, we study two general features of the eigenvectors, which
provide some evidence for good physical behavior of these solutions across
all affine-Virasoro space.

\section{An Origin for Braiding in Irrational CFT}

In rational CFT, braiding appears as a property of linear differential
equations, but, in the general CFT's of the Virasoro master equation, it is
unlikely that linear differential equations [20,7,14] extend beyond the coset
constructions. An important feature of the solution (8.2) is that it is based
on a linear (eigenvalue) problem, which, as we shall see, generates braiding
in the more general context.

To begin, we write the corresponding eigenvalue problem (8.1) in the more
flexible notation
$$  \eqalignno{ W_{qp} (u_0)  \bps_{p }^{(\n(u_0))} (u_0) & =
E_{\n(u_0)} (u_0) \bps_{ q }^{(\n(u_0))} (u_0) &(9.1a) \cr
 \ps_{q(\n(u_0))}(u_0)  W_{qp} (u_0 )  & =
E_{\n(u_0)} (u_0) \ps_{p (\n(u_0))} (u_0)  &(9.1b) \cr} $$
to facilitate comparison with the eigenvalue problem at $1-u_0$,
$$  \eqalignno{ W_{qp} (1-u_0)  \bps_{p }^{(\n(1-u_0))} (1-u_0) & =
E_{\n(1-u_0)} (1-u_0) \bps_{ q }^{(\n(1-u_0))} (1-u_0) &(9.2a) \cr
\ps_{q (\n(1-u_0))}(1-u_0)  W_{qp} (1-u_0 )  & =
E_{\n(1-u_0)} (1-u_0) \ps_{p (\n(1-u_0))} (1-u_0)  \;\;. \;\;\; &(9.2b) \cr} $$
In these forms, we avoid any labelling prejudice by allowing the
conformal structure index $\n$ to depend on $u_0$ or $1-u_0$.

The  eigenvalue problem at $1-u_0$ may be rewritten as
$$   W_{qp} (u_0) [ (-)^p P_{23} \bps_{p }^{(\n (1-u_0))} (1-u_0)]  =
E_{\n (1-u_0)}(1-u_0)[(-)^q P_{23}\bps_{q}^{(\n(1-u_0))} (1-u_0)]\eqno(9.3a) $$
$$  [ (-)^q \ps_{q (\n(1-u_0)) }(1-u_0) P_{23} ] W_{qp} (u_0 )
 =  E_{\n(1-u_0)} (1-u_0)[ (-)^p \ps_{p (\n (1-u_0))} (1-u_0) P_{23} ]
\eqno(9.3b)  $$
by using the crossing symmetry (4.8) of the connections. Comparing (9.3) to
the eigenvalue problem (9.1) at $u_0$, we learn first that the set of all
eigenvalues is closed under $u_0 \ra  1-u_0$
$$ \{ E_{\n(1-u_0)} (1 - u_0) \} = \{ E_{\n(u_0)} (u_0)  \} \eqno(9.4) $$
and we also learn that the connection eigenvectors enjoy the crossing symmetry
$$ (-)^p P_{23} \, \bps_{p \a} ^{(\n(1-u_0))} (1- u_0)  = \!\!
\sum_{(\m \,| E_{\m(u_0)} (u_0) = E_{\n(1-u_0)} (1- u_0) )}\! \!\!\!\!\!\!
\bps_{p \a}^{(\m (u_0)) } (u_0) \, \bX_{\m(u_0)} {}^{\n(1-u_0)} \eqno(9.5a)  $$
$$ (-)^p \ps_{p (\n(1-u_0))}^{\a} (1- u_0)  P_{23}  =  \!\!
\sum_{(\m \,| E_{\m(u_0)} (u_0) = E_{\n(1-u_0)} (1- u_0))} \! \!\!\!\!\!\!
 X_{\n(1-u_0)} {}^{\m(u_0)} \,\ps_{p (\m(u_0))}^{\a}(u_0)\;\;\;\;.\eqno(9.5b)
$$
The sums are over the $u_0$-eigenvectors with eigenvalue in the degenerate
subspace labelled by $ E_{\n(1-u_0)} (1- u_0)$, and $\bX, \,X$ are braid
matrices, to be determined.

The crossing symmetry (9.5) translates directly to the braiding of the
conformal eigenvectors
$$ P_{23}  \bps^{(\n(1-u_0))}_{\b} (1-u,1-u_0) =
\sum_{(\m \,| E_{\m(u_0)} (u_0) = E_{\n(1-u_0)} (1- u_0))} \!\!\!\!\!\!\!\!\!\!
  \bps^{(\m(u_0))}_{\b} (u,u_0) \, {\bX_{\m(u_0)}}{}^{\n(1-u_0)} \eqno(9.6a) $$
$$   \ps_{(\n(1-u_0))}^{\a} (1-u,1-u_0) P_{23} =
\sum_{(\m \,| E_{\m(u_0)} (u_0) = E_{\n(1-u_0)} (1- u_0))} \!\!\!\!\!\!\!\!\!\!
{X_{\n(1-u_0)}}^{\m(u_0)}  \,  \ps_{(\m(u_0))}^{\a} (u,u_0) \eqno(9.6b) $$
according to their definition in (8.2b,c). The braiding (9.6) of the conformal
eigenvectors, and the origin of the braiding in an eigenvalue problem, are
among the central results of this paper. It remains to study this braiding at
the level of conformal blocks, but, because the solution correctly includes the
correlators of the coset constructions, there can be little doubt that (9.6)
includes and generalizes the braiding of rational conformal field theory.

\section{The High-Level Correlators of Irrational CFT}
Beyond the coset constructions, it is unlikely that a closed form solution
can be obtained for the connection eigenvalue problem (8.1). On the other hand,
the problem is tractable by high-level expansion [12,14], which takes the
form of a degenerate perturbation theory.

For the expansion, we restrict ourselves to a fixed choice of external
representations $\T$ on
simple $g$, with conformal weights $\D (\T) = \cO(k^{-1})$ at high level. Then,
the invariant connections exhibit the form in (4.12b),
$$ W_{qp} = W_{q0} W_{0p} + V \;\;\;\;,\;\;\;\; V= \cO (k^{-2}) \eqno(10.1) $$
which  defines a Hamiltonian perturbation theory with leading-order Hamiltonian
$W_{q0} W_{0p}$ and perturbing potential $V$. The result at $V=0$ is exact to
all orders for $\tL =L_{g/h}$ and $L=L_h$, but, beyond the coset constructions,
the form of $V$ will generally violate the factorized form $W_{q0} W_{0p}$ of
the leading-order Hamiltonian.

The eigenvalue problem of the leading-order Hamiltonian
$$ \eqalignno{ W_{q0} (u_0 {)_{\a}}^{\g} [ \sum_{p} W_{0p} (u_0 {)_{\g}}^{\b}
 \bps_{p \b}^{(\n)} (u_0)] & = E_{\n} (u_0) \bps_{q \a}^{(\n)}(u_0) &(10.2a)
\cr
[ \sum_{q} \ps_{q(\n)}^{\b} (u_0)
W_{q0} (u_0 {)_{\b}}^{\g} ]  W_{0p} (u_0 {)_{\g}}^{\a}
 & = E_{\n} (u_0) \ps_{q (\n)}^{\a} (u_0)  &(10.2b) \cr} $$
is itself non-trivial, but, according to (8.2a), we need only those
eigenvectors with $E_{\n} \neq  0$, which are easily characterized by solving
(10.2) for the eigenvectors on the right. It follows that all the eigenvectors
with $E_{\n} \neq 0$ have the form
$$ \bps_{q \a}^{(\n)} (u_0) = W_{q0} (u_0 {)_{\a}}^{\b} \tf_{\b}^{(\n)} (u_0)
\;\;\;\;, \;\;\;\;  \ps_{p(\n)}^{\a} (u_0) = \f_{(\n)}^{\b} (u_0)
W_{0p} (u_0 {)_{\b}}^{\a}  \;\;\;\;,\;\;\;\;E_{\n} \neq 0 \eqno(10.3) $$
where $\tf$ and $\f$ are the (non-zero eigenvalue) eigenvectors of the
reduced eigenvalue problem
$$ M (u_0 {)_{\a}}^{\b} \tf_{\b}^{(\n)} (u_0)
= E_{\n} (u_0) \tf_{\a}^{(\n)} (u_0)  \;\;\;\;,\;\;\;\;
\f_{(\n)}^{\b} (u_0) M (u_0 {)_{\b}}^{\a} =  E_{\n} (u_0) \f_{(\n)}^{\a} (u_0)
\eqno(10.4a) $$
$$  M (u_0 {)_{\a}}^{\b} \equiv
 \sum_p ( W_{0p} (u_0)  W_{p0}  (u_0) {)_{\a}}^{\b}  \eqno(10.4b) $$
which is defined on the space $\T^1 \otimes \T^2 \otimes \T^3 \otimes \T^4$
of the representation matrices.

This result establishes the remarkable fact that only a finite number of
conformal structures
$$ \eqalignno{  \bY_{\n} (u,u_0) & = \sqrt{E_{\n} (u_0)} \,
Y_g^{\a}(u_0) \bps_{a}^{(\n)} (u,u_0) & \cr & = \sqrt{E_{\n} (u_0)} \,
Y_g^{\a} (u_0) \sum_{q=0}^{\infty} { (u-u_0)^q \over q!} \,W_{q0}(u_0)_{\a}
{}^{\b} \tf_{\b}^{(\n)} (u_0) + \cO (k^{-2}) \;\;\;\;\; \;\;\;\;\;\;
 &(10.5a) \cr
  Y_{\n}^{\a} (u,u_0) & = \sqrt{E_{\n} (u_0)} \, \ps^{a}_{(\n)} (u,u_0) & \cr
&= \sqrt{E_{\n} (u_0)} \, \f^{\b}_{(\n)} (u_0) \sum_{p=0}^{\infty}
W_{0p}(u_0)_{\b}{}^{\a} \,{ (u-u_0)^p \over p!} + \cO (k^{-2})  &(10.5b) \cr}
$$
$$ E_{\n} \neq 0 \;\;\;\;,\;\;\;\; \n = 0, 1,\ldots, D(\T)-1 \;\;\;\;,
\;\;\;\; D(\T) \leq \prod_{i=1}^4 {\rm dim}\, \T^i \eqno(10.5c) $$
contribute to the affine-Virasoro correlators at leading order. The counting is
a consequence of the factorized form $W_{q0} W_{0p} $ of the leading-order
Hamiltonian, a property which will  be lost at higher order for the generic
construction. In this case, higher-order perturbation theory will generate more
non-zero eigenvalue eigenvectors (from the infinite subspace $E_{\n} =0$ in
(10.2)), leading eventually to an infinite number of contributing conformal
structures for the generic affine-Virasoro correlator.

On the other hand, the form (10.1) at $V=0$, and hence the solution (10.5), is
exact to all orders for $\tL =L_{g/h} $ and $L=L_h$. With (8.5), this gives
$$ \bY_{\n}^{g/h} (u,u_0)  = Y_{g/h}^{\a} (u,u_0)  \sqrt{E_{\n} (u_0)} \,
\tf_{\a}^{(\n)} (u_0) \;\;\;\;,\;\;\;\; \n = 0, 1,\ldots, D(\T)-1 \eqno(10.6)
$$
for the coset constructions, in agreement with (8.6).

More generally, we may use the explicit form of the high-level connections
in (3.11), (4.9) and (4.11), for example,
$$ W_{q0} (u) = (-)^{q-1} (q-1) ! \, {\tilde{P}_{ab} \over k}
( \T_a^1 \T_b^2 u^{-q} + \T_a^1 \T_b^3 (u-1)^{-q} ) + \cO (k^{-2})
\;\;\;\;,\;\;q \geq 1 \eqno(10.7) $$
to sum the series in (10.5). The result summarizes the high-level
affine-Virasoro correlators
$$ \bY_{\n} (u,u_0) =  Y_{\tL}^{\a} (u,u_0) \, \sqrt{E_{\n} (u_0) } \,
 \tf_{\a}^{(\n)}(u_0) + \cO (k^{-2}) \eqno(10.8a) $$
$$   Y_{\tL}^{\a} (u,u_0)  =  Y_g^{\b} (u_0)  (
\one +  {\tilde{P}_{ab} \over k} \left[
\T_a^1  \T_b^2 \ln \left( {u \over u_0} \right)
+\T_a^1  \T_b^3 \ln \left( {1-u \over 1-u_0} \right) \right]
 )_{\b}{}^{\a} + \cO (k^{-2})     \eqno(10.8b) $$
$$ \tL^{ab} = { \tilde{P}^{ab} \over 2k} + \cO ( k^{-2}) \eqno(10.8c) $$
for all affine-Virasoro constructions $\tL$ on simple $g$.
The form (10.8) is one of the central results of this paper.

As seen above for the coset constructions, the high-level conformal structures
(10.8) are degenerate in that all the $u$ dependence of each structure is in
the same factor $Y_{\tL}^{\a}(u,u_0)$. We remark that the factor
$Y_{\tL}^{\a}(u,u_0)$ is the $n=4$ invariant form of the high-level
$n$-point correlators conjectured for all affine-Virasoro constructions in
eq.(14.3) of Ref.[14]. Since the coset correlators (10.6) are correctly
included in the exact solution, the form (10.8) with
$\tilde{P} = \tilde{P}_{g/h} = P_g -P_h$  is correct for all
high-level coset constructions. More generally, we will argue below that
the form shows good physical behavior for all the constructions.

More precisely, we will find that the factor $Y_{\tL}^{\a}(u,u_0)$ shows
the correct $\tL^{ab}$-broken conformal weights, and hence the correct
singularities, for an affine-Virasoro correlator of four broken affine primary
fields. To see this, we first expand $Y_g (u_0)$,
$Y_g(u_0) \sum_{i=1}^4 \T_a^i =0$ in a basis of invariant tensors $v_4^{\a}$ of
  $\T^1 \oti \T^2 \oti \T^3 \oti \T^4$,
$$ Y_g^{\a} (u_0) = \sum_{r,\x,\x'} \F_g(r,\x,\x';u_0) \, v_4^{\a}(r,\x,\x')
\;\;\;\;,\;\;\;\; v_4^{\b} (r,\x,\x') \sum_{i=1}^4
(\T_a^i)_{\b}{}^{\a} =0 \;\;\;\;.\eqno(10.9)$$
The coefficients $\F_g$ are related to the affine-Sugawara conformal blocks at
$u=u_0$, whose precise form is not relevant in the present discussion. One
choice for $v_4^{\a}$ is the s-channel basis
$$ v_4^{\a_1 \a_1 \a_3 \a_4 } (r,\x,\x') =
 \sum_{\a_r \a_{\br}}  v_3^{\a_1\a_2 \a_{\br} }(\x) \,
 v_3^{\a_3\a_4 \a_r }(\x') \, \et_{\a_{\br} \a_r} \eqno(10.10a)$$
$$ v_3^{\a_i\a_j \a_{\br} }(\x) =
\sum_{\a_{r}} \cl{\a_i}{\a_j}{r(\x)}{i}{j}{\a_{r}} \et^{\a_{r}\a_{\br}}
\eqno(10.10b) $$
$$ v_3^{\b_i \b_j \b_r}(\x)  (\T_a^i + \T_a^j + \T_a^r)_{\b_i \b_j \b_r}
{}^{\a_i \a_j \a_r} =0 \eqno(10.10c) $$
where  $v_3^{\a}(\x)$ are the invariant tensors of
$\T^i \otimes \T^j \otimes \T^r$, $\T^r$ an irreducible representation of $g$.
In (10.10b), these tensors are given in terms of the inverse metric
$\eta^{\a_r \a_{\br}} $ on the carrier space of $\T^r$ and the Clebsch-Gordan
coefficients $(\cdots )$ for the decomposition
$\T^i \otimes \T^j = \oplus_r \T^r$. The  $\x$ label in $v_3^{\a}$ is needed
when a representation $\T^r$ appears more than once in the decomposition.

Physically, the argument $r$ in $v_4^{\a}(r,\x,\x')$ labels the irreps $\T^r$
of $g$ which appear in the s channel ($u \ra 0$) of the four-point correlators,
while $\x ,\x'$ distinguish the different couplings of the various copies of
$\T^r$. The basis (10.10) was obtained by  studying Haar integration over four
representations of $g$, and corresponding t- and u-channel bases are obtained
by permutations of $\a_1 \a_2 \a_3 \a_4$ in (10.10a).

Using (3.1) and (10.10c), we verify the exact relation
$$ v_3^{\b_1 \b_2 \a_r} (\x)
[2 \tL^{ab} \T_a^1  \T_b^2]_{\b_1 \b_2 }{}^{\a_1 \a_2}
= v_3^{\a_1 \a_2 \a_{r}} (\x)
 (\tD_{\a_r} (\T^r) - \tD_{\a_1}(\T^1) - \tD_{\a_2}(\T^2)) \eqno(10.11) $$
where $\{\tD_{\a} (\T)\}= {\rm diag}(\tL^{ab} \T_a \T_b)$ are the conformal
weights of the broken affine-primary states corresponding to the external
representations $\T^1,\,\T^2$ and the s channel representation $\T^r$.

Collecting these results, we find that
$$   Y_{\tL}^{\a} (u,u_0)  \smash{ \mathop{\simeq} \limits_{u \ra 0} }
   \sum_{{r,\x,\x' \atop \a_r \a_{\br} } } \F_g(r,\x,\x';u_0) \,
v_3^{\a_1 \a_2 \a_{\br}} (\x) \,
\left( { u \over u_0} \right)^{\tD_{\a_r} - \tD_{\a_1} - \tD_{\a_2}}
v_3^{\a_3 \a_4 \a_r}(\x') \,   \et_{\a_{\br}\a_r}   $$
$$ + \cO (k^{-2}) \eqno(10.12) $$
which shows the correct conformal weight factor
$( u /u_0)^{\tD_{\a_r} - \tD_{\a_1} - \tD_{\a_2}}$ for broken
affine primaries in the s channel. A similar analysis in the t channel
($u \ra 1$) shows the expected factor
$ ((1-u) /( 1 -u_0))^{\tD_{\a_{t}} - \tD_{\a_1} - \tD_{\a_3}}$, where
$t$ labels the broken affine primaries in the t channel.

{}From (10.12), we may also read the high-level fusion rules of the broken
affine modules: In rational and irrational conformal field theory, these
rules follow the Clebsch-Gordan coefficients (10.10b) of the representations.
We remind the reader that the coefficients are computed in the simultaneous
$L$-basis of the representations (see Section 3), where all the conformal
weight matrices are diagonal. The next step is to study the braiding of the
high-level conformal blocks of (10.8).

\section{Conclusions}
The affine-Virasoro Ward identities [14] are a system of non-linear
differential equations which describe the correlators of all affine-Virasoro
constructions, including rational and irrational conformal field theory. In
Ref.[14], we solved the Ward identities for the coset constructions, providing
a derivation of the coset blocks of Douglas [15]. In this paper, we solved for
the conformal correlators of the affine-Sugawara nests, and showed that
global solutions exist across all affine-Virasoro space, so long as a
generically-infinite number of conformal structures is allowed. This is in
agreement with intuitive notions about irrational conformal field theory.

We focused on a particular global solution which is based on a natural
eigenvalue problem in the system. This solution reproduces the correct coset
and nest correlators and exhibits a braiding for all affine-Virasoro
correlators
which includes and generalizes the braiding of rational conformal field theory.
The underlying mechanism of the braiding is the linearity of the eigenvalue
problem.

The solution also shows good physical behavior, at least at high level on
simple $g$, where we are able to see the high-level correlators and high-level
fusion rules of irrational conformal field theory.

In this first look at the correlators of irrational conformal field theory,
we have raised as many questions as we have answered. In particular, further
work is necessary to be certain that our particular solution is globally
physical at higher order and/or finite level.

We are least satisfied in our understanding of the multiplicity of solutions to
the system, which is associated to various factorization ans\"atze. In spite of
appearances, we have seen some evidence that these solutions are related
to each other, sometimes via irrelevant constants and sometimes via a change of
basis, as noted in Section 7. Alternately, it is possible that we lack a
boundary condition on the system, whose nature could be central in finding the
correct solution.

To supplement future discussion of these questions, we have included in
Appendix D the results of another natural factorization, in the symmetric
ansatz. Although the assignment of the Lie algebra indices is quite different
in this solution, it gives the same correct coset and nest correlators, and
the same high-level correlators for all affine-Virasoro constructions.

\section*{Acknowledgements}
We acknowledge helpful conversations with M. Douglas, G. Rivlis, A. Sevrin and
E. Wichmann. Special thanks are due to E. Kiritsis, who participated in the
early stages of this work and pointed out that our biconformal nest correlators
should factorize into tensor-product theories.

The work of MBH was supported in part by the Director, Office of
Energy Research, Office of High Energy and Nuclear Physics, Division of
High Energy Physics of the U.S. Department of Energy under Contract
DE-AC03-76SF00098 and in part by the National Science Foundation under
grant PHY90-21139. The work of NO was supported in part by European
Community research project SC1-CT91-0674.

\bigskip
\bigskip
\centerline{ \bf Appendix A: Second-order connections}
\bigskip
For use in the text, we give the known forms of the second-order ($q+p=2$)
affine-Virasoro connections [14]. For the $n$-point connections, we have
$$ W_{0,ij} = \pa_i W_{0,j} + \frac{1}{2} (W_{0,i},W_{0,j})_+ +  E_{0,ij}
\;\;\;,\;\;\; W_{ij,0} = \pa_i W_{j,0} + \frac{1}{2} (W_{i,0},W_{j,0})_+ +
E_{ij,0} \eqno(A.1a) $$
$$  W_{i,j} = W_{i,0} W_{0,j} + E_{i, j} \eqno(A.1b)$$
$$ E_{i, j} = - 2i L^{da}L^{e(b} {f_{de}}^{c)} \left\{
{ \T_c^j \T_b^j \T_a^i + \T_c^i \T_b^i \T_a^j\over z_{ij}^2}
- 2 \sum_{k \neq i,j}^n  {\T_c^k \T_b^i \T_a^j \over z_{ij} z_{ik} } \right\}
 \;\;\;\; ,\;\;i \neq j   \eqno(A.1c)$$
$$ E_{0,ij} = -\frac{1}{2} ( E_{i, j} + E_{ j, i} ) \;\;\;\;,
\;\;\;  E_{ij,0} = E_{0,ij} E_{ i, i} =-\sum_{j \neq i}^n E_{i ,j}\eqno(A.1d)$$
and the corresponding invariant second-order connections are
$$ W_{02} = \pa W_{01} + W_{01}^2 + E_{02} \;\;\;,
\;\;\;\;W_{20} = \pa W_{10} + W_{10}^2 + E_{20} \eqno(A.2a) $$
$$ W_{11} = W_{10} W_{01} - E_{02} = W_{01} W_{10} - E_{20} \eqno(A.2b) $$
$$ E_{02}  = -2i  L^{da}  L^{e(b} {f_{de}}^{c)} V_{abc}
\;\;\;\;,\;\;\;\; E_{20} = E_{02} \ve_{L \ra \tL}  \eqno(A.2c)$$
$$ V_{abc} =\! \frac{1}{u^2} [\T_a^1 \T_b^2 \T_c^2 + \T_a^2 \T_b^1 \T_c^1]
  + \frac{1}{(u-1)^2}[\T_a^1 \T_b^3 \T_c^3 + \T_a^3 \T_b^1 \T_c^1] +
\frac{2}{u(u-1)} \T_a^1 \T_b^2 \T_c^3 \;. \eqno(A.2d)$$
More generally, the invariant one-sided connections $W_{0p}$ can
be obtained from the four-point connections by iterating the general
$SL(2,\R) \times SL(2,\R)$ relation
$$ \eqalign{ W_{0p} (u)  =  {1 \over f_{p,p} (z)}  & \left(
W_{0, 1 \ldots 1} (z) + (-)^{p+1}
{ \Ga (2\D_1 + p )  \over \Ga (2 \D_1) } {1 \over z_{14}^p }
- \sum_{s=1}^{p-1} f_{p,s}(z) W_{0s} (u) \right. \cr & \left.
- \sum_{r=1}^{p-1} \ch{p}{r} (-1)^r {\Ga (2\D_1 + r )  \over \Ga (2 \D_1) }
{1 \over z_{14}^r } \sum_{s=0}^{p-r} f_{p-r,s}(z) W_{0s}(u) \right) \cr}
\eqno(A.3a)$$
$$ \pa_1^p = \left( {\pa u \over \pa z_1} \pa_u \right)^p =
\sum_{s=1}^p f_{p,s} (z) \pa_u^s \eqno(A.3b) $$
and using the global Ward identity to eliminate the fourth representation
$\T^4$. Then, the mixed invariant connections $W_{qp}$ can be obtained from
$W_{0p}$ by using the solution (4.1c) of the invariant consistency relations.

\bigskip
\bigskip
\centerline{\bf Appendix B: General A-S nest connections}
\bigskip

The biconformal correlators (5.19) and (5.20) of the affine-Sugawara
(A-S) nests were obtained by continuing the iteration schematized in (5.18).
In this appendix, we use the general principles of the iteration to prove that
these results are correct.

By differentation of the invariant biconformal correlators (5.19) and
comparison with (3.10), we obtain the invariant connections of the general nest
$$ \eqalign{
 W_{q p} [\tL= L_{g/h_1/\ldots /h_{2n+1}},L= & L_{ h_1/\ldots /h_{2n+1}}]  \cr
= Y_g^{-1} \left\{ \bpa^q \pa^p [ Y_{g/h_1} (\bu) Y_{h_1/h_2} \right. &(u)
 \left. Y_{h_{2}/h_{3}} (\bu) \cdots
 Y_{h_{2n}/h_{2n+1}}(\bu) Y_{h_{2n+1}} (u) ] \right\}
\ve_{\bu =u} \cr
=  Y_g^{-1} \sum_{j_1=0}^q \sum_{j_2=0}^{j_1}  \cdots \sum_{j_n=0}^{j_{n-1}}
 & \sum_{i_1=0}^p  \sum_{i_2=0}^{i_1}  \cdots \sum_{i_n=0}^{i_{n-1}}
 \ch{q}{j_1}  \!\! \ch{p}{i_1} \prod_{k=2}^n \ch{j_{k-1}}{j_k} \!\!
\ch{i_{k-1}}{i_k} \cr
 \times
\pa^{q-j_1} Y_{g/h_1} \pa^{p-i_1} & Y_{h_1/h_2}  \pa^{j_1-j_2} Y_{h_2/h_3}
\pa^{i_1-i_2} Y_{h_3/h_4} \cdots \cr
\times \pa^{j_{n-1}-j_n} & Y_{h_{2n-2}/h_{2n-1}}
   \pa^{i_{n-1}-i_n} Y_{h_{2n-1}/h_{2n}} \pa^{j_n} Y_{h_{2n}/h_{2n+1}}
\pa^{i_n} Y_{h_{2n+1}} \cr} \eqno(B.1a) $$
$$ \eqalign{
  W_{qp}[\tL=L_{g/h_1/\ldots /h_{2n}}, &L= L_{ h_1/\ldots / h_{2n}} ]\cr
= Y_g^{-1} \left\{ \bpa^q \pa^p [ Y_{g/h_1} (\bu) \right. & Y_{h_1/h_2}  (u)
  \left. Y_{h_{2}/h_{3}} (\bu) \cdots
 Y_{h_{2n-1}/h_{2n}}(u) Y_{h_{2n}} (\bu) ] \right\} \ve_{\bu =u} \cr
= Y_g^{-1}  \sum_{j_1=0}^q \sum_{j_2=0}^{j_1}  \cdots & \sum_{j_n=0}^{j_{n-1}}
  \sum_{i_1=0}^p   \sum_{i_2=0}^{i_1} \cdots \sum_{i_{n-1}=0}^{i_{n-2}}
 \ch{q}{j_1} \!\! \ch{p}{i_1} \!\!\ch{j_{n-1}}{j_n} \prod_{k=2}^{n-1}
\ch{j_{k-1}}{j_k} \!\! \ch{i_{k-1}}{i_{k}} \cr
\times
\pa^{q-j_1} Y_{g/h_1} \pa^{p-i_1} & Y_{h_1/h_2}  \pa^{j_1-j_2} Y_{h_2/h_3}
\pa^{i_1-i_2} Y_{h_3/h_4} \cdots \cr
\times
\pa^{j_{n-1}-j_n} & Y_{h_{2n-2}/h_{2n-1}}
   \pa^{i_{n-1}-i_n} Y_{h_{2n-1}/h_{2n}} \pa^{j_n} Y_{h_{2n}} \;\;\;\;\;.
\cr}\eqno(B.1b) $$
These connections are guaranteed to  satisfy the consistency relations (3.12b),
which are the integrability conditions for the existence of the biconformal
correlators.

Then, we need only check the embedding relations
$$ W_{0 p} (L=  L_{h_1/\ldots  / h_{2n+1} } )
= W_{p 0} (\tL = L_{ h_1/\ldots  / h_{2n+1} } ) \eqno(B.2a) $$
$$ W_{0 p} (L= L_{h_1/\ldots  / h_{2n} } )
= W_{p 0} (\tL = L_{ h_1/\ldots  / h_{2n}} ) \eqno(B.2b) $$
which are verified from (B.1) by choosing first $q=0$ and then $p=0$, followed
by the appropriate renaming of groups.

As another check on these results, note that the form (B.1b) for the even nests
can be obtained from the form (B.1a) for the odd  nests by setting
$h_{2n+1} =0$ and $Y_{h_{2n+1}} = 1 $.

An alternate expression for the invariant nest connections
$$ \eqalign{
W_{q p}  & [\tL=  L_{g/h_1/ \ldots /h_{2n+1}}, L=L_{ h_1/\ldots/ h_{2n+1}}]
=\cr
\sum_{j_1=0}^q  &\sum_{j_2=0}^{j_1}  \cdots \sum_{j_n=0}^{j_{n-1}}
 \sum_{i_1=0}^p \sum_{i_2=0}^{i_1} \cdots \sum_{i_n=0}^{i_{n-1}}
 \ch{q}{j_1}  \ch{p}{i_1} \prod_{k=2}^n \ch{j_{k-1}}{j_k} \ch{i_{k-1}}{i_k} \cr
 \times &
W_{0,q-j_1}^{g/h_1} W_{0,p-i_1}^{h_1/h_2}  W_{0,j_1-j_2}^{h_2/h_3}
W_{0,i_1-i_2}^{h_3/h_4}  \cdots  W_{0,j_{n-1}-j_n}^{h_{2n-2}/h_{2n-1}}
  W_{0,i_{n-1}-i_n}^{h_{2n-1}/h_{2n}} W_{0j_n}^{h_{2n}/h_{2n+1}}
W_{0i_n}^{h_{2n+1}} \cr} \eqno(B.3) $$
follows with (5.8), and the corresponding result for even nests is obtained
from (B.3) with $h_{2n+1}=0$, $Y_{h_{2n+1}}=1$ and
$W_{0 i_n}^{h_{2n+1}} =\d_{i_n,0} $. This form of the connections shows that
they are independent of the reference point $u_0$ of the evolution operators,
and it is also the most convenient form to check against the known forms of the
first and second-order connections in (3.11) and (A.2). After some algebra, we
find that they are in complete agreement.

We turn now to the general $n$-point  biconformal nest correlators in (5.20).
To check this result, we need the general $n$-point nest connections
$$ \eqalign{  W_{j_1 \ldots j_q} & {}_{,i_1 \ldots i_p}
[\tL= L_{g/h_1/\ldots /h_{2n+1}},  L=L_{ h_1/\ldots /h_{2n+1} }]  =\cr
 A_g^{-1} & \sum_{\P(j_1 \ldots j_q)} {1 \over q!} \sum_{\P(i_1 \ldots i_p)}
{1 \over p!}  \sum_{l_1=0}^q  \sum_{l_2=0}^{l_1} \cdots \sum_{l_n=0}^{l_{n-1}}
 \sum_{k_1=0}^p \sum_{k_2=0}^{k_1} \cdots \sum_{k_{n}=0}^{k_{n-1}} \cr
& \times
\ch{q}{l_1} \ch{p}{k_1} \prod_{r=2}^n \ch{l_{r-1}}{l_r}   \ch{k_{r-1}}{k_r}
\left[ ( \prod_{\m_1 = 1}^{q -l_1} \pa_{j_{\m_1}} )  A_{g/h_1} \right]
\left[ (\prod_{\n_1 = 1}^{p -k_1} \pa_{i_{\n_1}} )   A_{h_1/h_2} \right] \cr
& \times
\left[ ( \! \prod_{{\m_2 = \atop q-l_1+1}}^{q -l_2} \!\!\! \pa_{j_{\m_2}}
)  A_{h_2/h_3} \right]
\left[ ( \!\prod_{{\n_2 = \atop p-k_1+1}}^{p -k_2} \!\!\! \pa_{i_{\n_2}}
)   A_{h_3/h_4} \right] \cdots
\left[ ( \! \prod_{{\m_n = \atop q-l_{n-1}+1}}^{q -l_n} \!\!\!\!
\pa_{j_{\m_n}} )  A_{h_{2n-2}/h_{2n-1}} \right] \cr
& \times
\left[ ( \! \prod_{{\n_n = \atop p-k_{n-1}+1}}^{p -k_n}\!\!\!\!
\pa_{i_{\n_n}} ) A_{h_{2n-1}/h_{2n}} \right] \!\!
\left[ ( \!\prod_{{\m_{n+1} = \atop q-l_{n}+1}}^{q} \!\!\! \pa_{j_{\m_{n+1}}}
)  A_{h_{2n}/h_{2n+1}} \right] \!\!
\left[ ( \! \prod_{{\n_{n+1} = \atop p-k_{n}+1}}^{p} \!\!\! \pa_{i_{\n_{n+1}}}
) A_{h_{2n+1}} \right] \cr}\eqno(B.4a)$$
$$ \eqalign{
= & \sum_{\P(j_1 \ldots j_q)} {1 \over q!} \sum_{\P(i_1 \ldots i_p)}
{1 \over p!}  \sum_{l_1=0}^q \sum_{l_2=0}^{l_1} \cdots \sum_{l_n=0}^{l_{n-1}}
 \sum_{k_1=0}^p \sum_{k_2=0}^{k_1} \cdots \sum_{k_{n}=0}^{k_{n-1}}
 \ch{q}{l_1} \ch{p}{k_1} \cr
&  \times
\prod_{r=2}^n \ch{l_{r-1}}{l_r} \!\!  \ch{k_{r-1}}{k_r}
  W_{0,j_1 \cdots j_{q-l_1} }^{g/h_1}  W_{0,i_1 \cdots i_{p-k_1}}^{h_1/h_2}
  W_{0,j_{q-l_1+1} \cdots j_{q-l_2}}^{h_2/h_3}
W_{0,i_{p-k_1+1} \cdots i_{p-k_2}}^{h_3/h_4}  \cr
& \;\;\; \cdots
 W_{0,j_{q-l_{n-1}+1} \cdots j_{q-l_n}}^{h_{2n-2}/h_{2n-1}}
W_{0,i_{p-k_{n-1}+1} \cdots i_{p-k_n}}^{h_{2n-1}/h_{2n}}
W_{0,j_{q-l_n+1} \cdots j_{q}}^{h_{2n}/h_{2n+1}}
W_{0,i_{p-k_n+1} \cdots i_{p}}^{h_{2n+1}}   \cr} \eqno(B.4b)$$
which are obtained by differentation of (5.20) and comparison with (3.5). Here,
$W_{0,i_1 \ldots i_p}^{g/h} =W_{0,i_1 \ldots i_p}(L_{g/h})$ and
$W_{0,i_1 \ldots i_p}^{h} =W_{0,i_1 \ldots i_p}(L_{h})$ are  the one-sided
$n$-point connections of $g/h$ and $h$, and the results for the even nests
$g/h_1 / \ldots /h_{2n}$ can be obtained from (B.4) by setting $h_{2n+1}=0$,
$A_{h_{2n+1}} = 1$ and $W_{i_{p-k_n +1} \ldots i_p}^{h_{2n+1}} \ra \d_{k_n,0}$.
The general nest connections are independent of the reference point $z_0$
of the evolution operators, and satisfy the consistency relations (3.12a),
which are the integrability conditions for the biconformal correlators.
Finally, the  correct embedding relations
$$ \eqalignno{  W_{0, i_1 \ldots i_p} (L=  L_{h_1/\ldots  / h_{2n+1} } )
& = W_{i_1 \ldots i_p, 0} (\tL = L_{ h_1/\ldots  / h_{2n+1} } ) &(B.5a) \cr
W_{0 ,i_1 \ldots i_p} (L= L_{h_1/\ldots  / h_{2n} } )
&= W_{i_1 \ldots i_p, 0} (\tL = L_{ h_1/\ldots  / h_{2n}} ) &(B.5b)\cr} $$
are  verified from (B.4) and the corresponding form for the even nests,  which
completes the check of (5.20). We have also checked (B.4) against the known
forms of the first and second-order connections in (3.8) and (A.1).

\bigskip
\bigskip
\centerline{\bf Appendix C: Conformal blocks of the A-S nests}
\bigskip
Following the change of basis given for the coset correlators in Ref.[14], we
discuss the affine-Sugawara (A-S) nests at the level of conformal blocks.

For four representations $\T^i_a,\;i=1,\ldots,4$ of $g$ and the subgroup
sequence $g \supset h_1 \supset \ldots \supset h_n$, we introduce the
$h_j$-invariant tensors $v_{m_j}^{\a} (h_j)$,
$$ v_{m_j}^{\b} (h_j)\sum_{i=1}^4 (\T_{a}^i)_{\b}{}^{\a} = 0\;\;\;\;,\;\;\;\;
a \in h_j   \eqno(C.1a) $$
$$ \{ v_{m_j}^{\a} (h_j) \} \subset \{v_{m_{j+1}}^{\a} (h_{j+1}) \}
\;\;\;\;,\;\;\;\; j=0,1, \ldots, n-1  \;\;\;\;\;\;. \eqno(C.1b)$$
Here we have introduced $h_0 \equiv g$ for uniformity and the tensors are
chosen to satisfy $v_{m_j}^{\a} (h_j)= v_{m_{j}}^{\a} (h_{j+1}),\;\,
\{m_j \} \subset \{m_{j+1} \} $. Using global Ward identities, we may then
expand the operators in (5.19) as [14]
$$ Y_{h_0/h_1}^{\a}(u,u_0) =  Y_{h_0/h_1}^{m_1} (u,u_0) v_{m_1}^{\a} (h_1)
=  d^{r_0} \C_{h_0/h_1} (u)_{r_0}{}^{r_1} \F_{h_1} (u_0)_{r_1}{}^{m_1}
v_{m_1}^{\a} (h_1) \eqno(C.2a) $$
$$  {Y_{h_j}(u,u_0)_{m_j}}^{\a} \equiv v_{m_j}^{\b} (h_j)
{Y_{h_j}(u,u_0)_{\b}}^{\a} = \F_{h_j}^{-1} (u_0)_{m_j}{}^{r_j}
\F_{h_j} (u)_{r_j}{}^{n_j}  v_{n_j}^{\a}(h_j) \eqno(C.2b) $$
$$ v_{m_j}^{\b} (h_{j})Y_{h_j/h_{j+1}}(u,u_0)_{\b}{}^{\a}\equiv Y_{h_j/h_{j+1}}
(u,u_0)_{m_j}{}^{m_{j+1}} v_{m_{j+1}}^{\a} (h_{j+1}) \eqno(C.2c) $$
$$ Y_{h_j/h_{j+1} } (u,u_0)_{m_j}{}^{m_{j+1}} =
\F_{h_j}^{-1} (u_0)_{m_j}{}^{r_j} \C_{h_j/h_{j+1}}(u)_{r_j}{}^{r_{j+1}}
   \F_{h_{j+1}} (u_0)_{r_{j+1}}{}^{m_{j+1}}  \eqno(C.2d)$$
$$ \C_{h_j/h_{j+1}}(u)_{r_j}{}^{r_{j+1}} =
\F_{h_j}(u)_{r_j}{}^{m_j}  \F_{h_{j+1}}^{-1}(u)_{m_j}{}^{r_{j+1}} \;\;\;\;.
\eqno(C.2e)  $$
Here, $d^{r_0}$ are constants, $\F_{h_j}$ are the conformal blocks of $h_j$,
chosen so that the left indices $r_j$ label the blocks by $h_j$ representations
in the s channel ($u \ra 0$), and $\C$ are the coset blocks [15,14].

Using (C.2), the biconformal correlators of the nests (5.19) factorize
as follows,
$$  \eqalign{ Y^{\a} & (\bu,u)
[\tL = L_{g/h_1 / \ldots /h_{2n+1}}, L = L_{h_1 / \ldots /h_{2n+1}}]  \cr
 = & Y_{g/h_1}^{m_1} (\bu,u_0) Y_{h_1/h_2} (u,u_0)_{m_1}{}^{m_2}
    \cdots Y_{h_{2n}/h_{2n+1}}(\bu,u_0)_{m_{2n}}{}^{m_{2n+1}}
Y_{h_{2n+1}}(u,u_0)_{m_{2n+1}}{}^{\a}  \cr
 = & \bY_{\n}(\bu,u_0) [\tL = L_{g/h_1 / \ldots /h_{2n+1}}] \,
 Y_{\n}^{\a}(u,u_0) [ L = L_{h_1 / \ldots /h_{2n+1}}] \;,\;
\n \equiv (m_1,\ldots,m_{2n+1})  \cr} \eqno(C.3a)$$
$$  \eqalign{ Y^{\a} & (\bu,u)
[\tL = L_{g/h_1 / \ldots /h_{2n}}, L = L_{h_1 / \ldots /h_{2n}}]  \cr
 = & Y_{g/h_1}^{m_1} (\bu,u_0) Y_{h_1/h_2} (u,u_0)_{m_1}{}^{m_2}
  \cdots Y_{h_{2n-1}/h_{2n}}(u,u_0)_{m_{2n-1}}{}^{m_{2n}}
Y_{h_{2n}}(\bu,u_0)_{m_{2n}}{}^{\a}  \cr
 = & \bY_{\m}^{\a} (\bu,u_0) [\tL = L_{g/h_1 / \ldots /h_{2n}}] \,
Y_{\m}(u,u_0) [ L = L_{h_1 / \ldots /h_{2n}}] \;\;\;,\;\;\;
\m \equiv (m_1,\ldots,m_{2n})  \cr}  \eqno(C.3b)$$
where the factorized correlators of the $\tL$ theories are
$$ \eqalignno{
\bY_{\n}(\bu,u_0) & [\tL = L_{g/h_1 / \ldots /h_{2n+1}}] & \cr
& =  Y_{g/h_1}^{m_1}(\bu,u_0) \prod_{j=1}^n Y_{h_{2j}/h_{2j+1}}
(\bu,u_0)_{m_{2j}}{}^{m_{2j+1}}  &(C.4a) \cr
 & = f_{\n}(u_0)^{r_0r_2  \ldots r_{2n}}_{r_1 r_3  \ldots r_{2n+1}}
 \, \N_{g/h_1/ \ldots/h_{2n+1}}
(\bu)_{r_0 r_2  \ldots r_{2n}}^{r_1 r_3  \ldots r_{2n+1}} \;\;\;\;\;\;\;\;
&(C.4b) \cr} $$
$$ \eqalignno{
\bY^{\a}_{\m}(\bu,u_0) & [ \tL = L_{g/h_1 / \ldots /h_{2n}}] & \cr
&= Y_{g/h_1}^{m_1}(\bu,u_0) \prod_{j=1}^{n-1} Y_{h_{2j}/h_{2j+1}}
(\bu,u_0)_{m_{2j}}{}^{m_{2j+1}}  Y_{h_{2n}}(\bu,u_0)_{m_{2n}}{}^{\a}
&(C.4c) \cr
& = f_{\m}^{\a}
(u_0)^{r_0r_2 \ldots r_{2n-2} r_{2n}}_{r_1 r_3 \ldots r_{2n-1} m'{}_{2n}}
  \, \N_{g/h_1/ \ldots/h_{2n}}
(\bu)_{r_0r_2  \ldots r_{2n-2} r_{2n}}^{r_1 r_3  \ldots r_{2n-1} m'{}_{2n}}
\;\;\;\;. &(C.4d) \cr} $$
The constant factors $f(u_0)$ in (C.4) are
$$ f_{\n}(u_0)^{r_0r_2  \ldots r_{2n}}_{r_1 r_3  \ldots r_{2n+1}}
= d^{r_0} \prod_{j=1}^n \F_{h_{2j}}^{-1}(u_0)_{m_{2j}}{}^{r_{2j}}
\prod_{k=0}^n \F_{h_{2k+1}}(u_0)_{r_{2k+1}} {}^{m_{2k+1}} \eqno(C.5a)$$
$$ f_{\m}^{\a}
(u_0)^{r_0r_2  \ldots r_{2n-2} r_{2n}}_{r_1 r_3  \ldots r_{2n-1} m'{}_{2n}}
= d^{r_0} \prod_{j=0}^{n-1} \F_{h_{2j}}^{-1}(u_0)_{m_{2j}}{}^{r_{2j}}
\F_{h_{2j+1}}(u_0)_{r_{2j+1}}{}^{m_{2j+1}}  v^{\a}_{m'{}_{2n}} (h_{2n})
\eqno(C.5b)$$
and the nest blocks $\N_{g/h_1 / \ldots / h_n}$ of the conformal field
theories  $L_{g/h_1 / \ldots / h_n}$ are given by
$$ \eqalignno{ \N_{g/h_1/ \ldots/h_{2n+1}}(\bu)_{r_0r_2 \ \ldots r_{2n}}
^{r_1 r_3 \ \ldots r_{2n+1}}
& = \prod_{j=0}^n \C_{h_{2j}/h_{2j+1}} (\bu)_{r_{2j}}{}^{r_{2j+1}}
 \;\;\;\;,\;\;h_0=g &(C.6a) \cr
 \N_{g/h_1/ \ldots/h_{2n}}(\bu)_{r_0r_2 \ \ldots r_{2n-2} r_{2n}}
^{r_1 r_3 \ \ldots r_{2n-1} m_{2n}}
& = \prod_{j=0}^{n-1} \C_{h_{2j}/h_{2j+1}} (\bu)_{r_{2j}}{}^{r_{2j+1}}
\F_{h_{2n}}(\bu)_{r_{2n}}{}^{m_{2n}} \;\;.\;\;\;\; &(C.6b) \cr} $$
Each of the nest blocks $\N$ is a tensor product of blocks, as expected.

As a consistency check on the factorization, it is not difficult to check that
the correlators $Y(L)$ in (C.3) give the blocks $\N_{h_1 / \ldots / h_{2n+1}}$
and $\N_{h_1 / \ldots / h_{2n}}$, obtained from (C.6) by renaming groups.
Note also that eqs.(C.4a,c) are the explicit forms of the tensor products
(5.21).

As a concrete example on $g\supset h_1 \supset \ldots \supset h_n$, we choose
the subgroup nest
$$  g_x =(h_0)_x = SU(N)_{x_0} \times SU(N)_{x_1} \times \ldots \times
SU(N)_{x_{n}} \eqno(C.7a) $$
$$ h_j = SU(N)_{y_j} \, \times_{l=j+1}^{n} SU(N)_{x_l} \;\;\;\;,
\;\;\;\; y_j \equiv \sum_{k=0}^j x_k  \eqno(C.7b)$$
and the integrable representations of $g$ as
$ \T^1= \T^4 =(\T_{(N)},0,\ldots,0),\;
\T^2 = \T^3 =(\bar{\T}_{(N)},0,\ldots,0)$  in the $N$ and $\bar{N}$ of
$SU(N)_{x_0}$. Then (C.6) gives the nest blocks
$$ \eqalignno{
(\N_{g/h_1/ \ldots/h_{2n+1}})_{r_0r_2 \ \ldots r_{2n}}
^{r_1 r_3 \ \ldots r_{2n+1}}
&= \prod_{j=0}^n (\C_{y_{2j},y_{2j+1}} )_{r_{2j}}{}^{r_{2j+1}}
\;\;\;\;,\;\; h_0 = g &(C.8a) \cr
 (\N_{g/h_1/ \ldots/h_{2n}})_{r_0r_2 \ \ldots r_{2n-2} r_{2n}}
^{r_1 r_3 \ \ldots r_{2n-1} m_{2n}}
& = \prod_{j=0}^{n-1} (\C_{y_{2j},y_{2j+1}})_{r_{2j}}{}^{r_{2j+1}}
(\F_{y_{2n}})_{r_{2n}}{}^{m_{2n}}  &(C.8b) \cr} $$
where the coset blocks $\C$ and the $SU(N)$ blocks $\F$ are 2$\times$2
matrices,
$$ (\C_{y_j,y_{j+1}})_{r_j}{}^{r_{j+1}} =
 (\F_{y_j})_{r_j}{}^{m_j} (\F_{y_{j+1}}^{-1})_{m_j}{}^{r_{j+1}} \;\;\;,
\;\; r_j= V,A \;\;,\;\;m_j =1,2 \;\;\;\;\forall \;j  \eqno(C.9a) $$
$$ \F_{y} = \left( \matrix{ {(\F_{y})_V}^1 &  {(\F_{y})_V}^2 \cr
  {(\F_{y_x})_A}^1 &  {(\F_{y})_A}^2 } \right) \,,\, \F_{y}^{-1}  =
 -\frac{1}{N} (u(1-u))^{4\D_{y} - \D_{y}^A}
\left( \matrix{ {(\F_{y})_A}^2 &  -{(\F_{y})_V}^2 \cr
               - {(\F_{y})_A}^1 &  {(\F_{y})_V}^1 } \right)\eqno(C.9b)$$
$$ \D_y = { N^2 -1 \over 2N(y+N) }
\;\;\;,\;\;\;\; \D_{y}^A = {N \over y + N} \;\;\;,\;\; \;
\l_{y} ={ 1 \over y+N}  \eqno(C.9c)$$
$$  \eqalign{ {(\F_{y}(u))_V}^1 & = u^{-2 \D_{y}} (1-u)^{\D_{y}^A - 2\D_{y}}
F(\l_y,-\l_y,1-N\l_y;u) \cr
 {(\F_{y}(u))_A}^1  &= u^{ \D_{y}^A - 2\D_{y}}
(1-u)^{\D_{y}^A - 2\D_{y}} F((N-1)\l_y,(N+1)\l_y,1+ N\l_y;u) \cr
{(\F_{y}(u))_V}^2  & = \frac{1}{y} u^{-2 \D_{y} +1}
(1-u)^{\D_{y}^A - 2\D_{y}}  F(1+\l_y,1-\l_y,2-N\l_y;u) \cr
 {(\F_{y}(u))_A}^2  & =-N u^{ \D_{y}^A - 2\D_{y}}
(1-u)^{\D_{y}^A - 2\D_{y} }
F((N-1)\l_y,(N+1)\l_y,N\l_y;u) \;\;.\cr} \eqno(C.9d)$$
Here $V$ and $A$ label the vacuum and adjoint blocks [7] in the $u\ra 0$
channel
and $F$ is the hypergeometric function. In these examples, the number of nest
blocks $\N_{g/h_1/ \ldots/h_{n}}$ is $2^{n+1}$. It is interesting to note that,
for a fixed choice of external representations, the number of nest blocks grows
with the nest depth $n$. In this sense, the nests may be considered as a
prelude
to irrational conformal field theory, where a generically-infinite number of
blocks is expected.

We have made a spot check of the s-channel singularities of the nest blocks in
(C.8), using the known behavior of the subgroup and coset blocks [14] as
$u\ra 0$. Approximately half of these intermediate states are immediately
identifiable as broken affine primary states (with conformal weights
$\D = \sum_{j=0}^n (-)^j \D_{h_j}$) and the other states are presumably broken
affine secondary.

Finally, the crossing-symmetric non-chiral correlators of these nests
$$ \eqalignno{
\Y_{g/h_1/ \ldots/h_{2n+1}} (u,u^*)
 & = \prod_{j=0}^n \Y_{h_{2j}/h_{2j+1}} (u,u^*)
\;\;\;\;,\;\;h_0=g &(C.10a) \cr
 \Y_{g/h_1/ \ldots/h_{2n}} (u,u^*) &=\prod_{j=0}^{n-1}
\Y_{h_{2j}/h_{2j+1}} (u,u^*) \Y_{h_{2n}}(u,u^*) &(C.10b) \cr} $$
$$ \eqalign{
\Y_{h_{j}}  (u,u^*) = & \F_{y_j} (u)_V{}^1 \F_{y_j} (u^*)_V{}^1
 + \F_{y_j} (u)_V{}^2 \F_{y_j} (u^*)_V{}^2 \cr
 &  + f(\l_{y_j})^{-1} [\F_{y_j} (u)_A{}^1 \F_{y_j} (u^*)_A{}^1
 + \F_{y_j} (u)_A{}^2 \F_{y_j} (u^*)_A{}^2 ] \cr} \eqno(C.10c)$$
 $$   \Y_{h_{j}/h_{j+1}} (u,u^*) =
\C_{y_j,y_{j+1}} (u)_V{}^V \C_{y_j,y_{j+1}} (u^*)_V{}^V
 +  f(\l_{y_{j+1}}) \C_{y_j,u_{j+1}} (u)_V{}^A \C_{y_j,y_{j+1}} (u^*)_V{}^A $$
 $$ + f(\l_{y_j})^{-1} [\C_{y_j,y_{j+1}} (u)_A{}^V \F_{y_j,y_{j+1}} (u^*)_A{}^V
 + f(\l_{y_{j+1}}) \C_{y_j,y_{j+1}} (u)_A{}^A
\C_{y_j,y_{j+1}} (u^*)_A{}^A ] \eqno(C.10d)$$
 $$ f(\l) \equiv  N^2 \left( \Ga (N\l) \over \Ga (1-N\l) \right)^2
{\Ga(1-(N-1)\l)\Ga (1-(N+1) \l) \over \Ga((N-1)\l) \Ga((N+1)\l)}
\eqno(C.10e) $$
are nothing but the product of the crossing-symmetric non-chiral correlators of
the relevant subgroups [7] and cosets [14]. At level $x_0=1$, the number of
contributing nest blocks in (C.10) is $2^n$, which corresponds to the usual
consistent chiral truncation of the blocks of $SU(N)_1$.

\bigskip
\bigskip
\centerline{\bf Appendix D: Symmetric factorization}
\bigskip

In this appendix, we discuss a second natural factorization of the invariant
biconformal correlators, in the symmetric ansatz (7.2d). This gives a second
global solution, whose details apparently differ from the solution of the text.
Nevertheless, we find that this solution gives the same correct coset and nest
correlators, and the same high-level affine-Virasoro correlators found in
Section 10.

At each fixed choice of the Lie algebra index $\a=(\a_1\a_2\a_3\a_4)$,
the matrix $C_{qp}^{\a} = Y_g^{\b} (u_0) W_{qp} (u_0 {)_{\b}}^{\a} $
defines an infinite-dimensional eigenvalue problem
$$ \sum_p  C_{qp}^{\a} \bps_{p(\n)}^{\a}(u_0) = E_{\n}^{\a}(u_0)
\bps_{q(\n)}^{\a}(u_0)  \eqno(D.1a) $$
$$ \sum_q \ps_{q(\n)}^{\a}(u_0) C_{qp}^{\a} = E_{\n}^{\a}(u_0)
\ps_{p(\n)}^{\a}(u_0) \eqno(D.1b) $$
$$  C_{qp}^{\a} =\sum_{\n=0}^{\infty}
 \bps_{q(\n)}^{\a} (u_0) E_{\n}^{\a}(u_0) \ps_{p(\n)}^{\a}(u_0)\;\;\;\;.
 \eqno(D.1c)$$
Then, the spectral resolution (D.1c) of $C_{qp}^{\a}$ gives the factorization
$$ \bY_{\n}^{\a} (u,u_0) = \sqrt{E_{\n}^{\a}(u_0)} \,\bps_{(\n)}^{\a}(u,u_0)
\;\;\;\;,\;\;\;\; \bps_{(\n)}^{\a}(u,u_0) \equiv \sum_{q=0}^{\infty}
{ (u-u_0)^q \over q!} \,\bps_{q(\n)}^{\a} (u_0) \eqno(D.2b)$$
$$ Y_{\n}^{\a} (u,u_0) = \sqrt{E_{\n}^{\a}(u_0)} \, \ps_{(\n)}^{\a}(u,u_0)
 \;\;\;\;,\;\;\;\; \ps_{(\n)}^{\a}(u,u_0) \equiv \sum_{p=0}^{\infty}
{ (u-u_0)^p \over p!} \,\ps_{p(\n)}^{\a} (u_0) \eqno(D.2c)$$
where the structures $\bps_{(\n)}^{\a}(u,u_0)$ and $\ps_{(\n)}^{\a}(u,u_0)$
are the conformal eigenvectors of the $\tL$ and $L$ theories respectively.

An elegant feature of the symmetric factorization is that the conformal
structures are symmetric under K-conjugation,
$$ \tL \lra L \;: \;\;\;\;\;\bY^{\a}_{\n}  \lra Y^{\a}_{\n}
\;\;\;\;. \eqno(D.3) $$
This symmetry, not shared by the solution of the text, follows from the
 K-conjugation covariance (4.5b) which implies that  $C_{qp} \lra C_{pq}$ and
hence $\bps \lra \ps$ when $\tL \lra L$.  On the other hand, the braiding of
the symmetric factorization is further complicated by the factor $Y_g(u_0)$ in
$C_{qp}$.

Corresponding to (8.3), an equivalent form of the factorized correlators
$$ \bY_{\n}^{\a} (u,u_0) = {1 \over \sqrt{E_{\n}^{\a}(u_0)} } \,
\sum_{p=0}^{\infty} \bps_{p(\n)}^{\a} (u_0) \pa_{u_0}^p \fb_0^{\a} (u,u_0)
\eqno(D.4a) $$
$$ Y_{\n}^{\a} (u,u_0) = {1 \over \sqrt{E_{\n}^{\a}(u_0)} } \,
\sum_{q=0}^{\infty} \ps_{q(\n)}^{\a} (u_0) \pa_{u_0}^q f_0^{\a} (u,u_0)
\eqno(D.4b) $$
is obtained from the eigenvalue problem (D.1) when $E_{\n} \neq 0$. The basic
structures  $\fb_0^{\a} $ and $f_0^{\a} = Y_g^{\b} (u_0) (f_0 {)_{\b}}^{\a}$
are defined in eqs.(8.3c,d).

As an application of (D.4), we give the results for the coset constructions and
the first non-trivial affine-Sugawara nests
$$ \eqalignno{
 (\bY^{g/h})_{\n}^{\a} (u,u_0) & = Y_{g/h}^{\b} (u,u_0) d_{\b (\n)}^{\a} (u_0)
 &(D.5a) \cr
d_{\b (\n)}^{\a} (u_0) & \equiv {1 \over \sqrt{E_{\n}^{\a}(u_0)} } \,
\sum_{p=0}^{\infty}  W_{0p}^h (u_0)_{\b}{}^{\a} \bps_{p(\n)}^{\a} (u_0)
&(D.5b) \cr} $$
$$ \eqalignno{
(\bY^{g/h_1/h_2})_{\n}^{\a} (u,u_0)  &= Y_{g/h_1}^{m_1} (u,u_0)
Y_{h_2} (u,u_0)_{m_2}{}^{\a} D_{m_1(\n)}^{m_2\a} (u_0) &(D.5c) \cr
D_{m_1(\n)}^{m_2\a} (u_0) & \equiv {1 \over \sqrt{E_{\n}^{\a}(u_0)} } \,
  \sum_{p=0}^{\infty} W_{0p}^{h_1/h_2} (u_0)_{m_1}{}^{m_2}  \,
\bps_{p(\n)}^{\a} (u_0) &(D.5d) \cr} $$
which correspond to  $\tL=L_{g/h}$ and $\tL=L_{g/h_1/h_2}$ respectively.
The symbols in (D.5c,d) with $m_1$ and/or $m_2$ indices are defined in
eqs.(5.16)
and (8.7d). Although the constants in (D.5) differ from those in the
corresponding results (8.6) and (8.7), the same correct $u$ dependence is
obtained for these conformal correlators. Simlilarly, the global solution (D.2)
or (D.4) gives the correct $u$ dependence for the conformal correlators of the
higher nests.

Finally, we follow the steps in Section 10 to obtain the high-level correlators
in the symmetric ansatz. As in the vector ansatz, only a finite number of
$E_{\n} \neq 0$ eigenvectors are found at leading order
$$  \eqalign{ \bps_{q(\n)}^{\a} (u_0) & = (Y_g(u_0) W_{q0}(u_0))^{\b}
 \tf_{\b(\n)}^{\a} (u_0) + \cO (k^{-2} ) \cr
 \ps_{p(\n)}^{ \a} (u_0) &  = (Y_g(u_0) W_{0p}(u_0))^{\b}
 \f_{\b (\n)}^{\a} (u_0) + \cO (k^{-2} ) \cr  }  \eqno(D.6a)$$
$$ E_{\n} \neq 0 \;\;\;\;,\;\;\;\; \n =0, 1, \ldots ,D_s (\T)-1
\;\;\;\;,\;\;\;\; D_s (\T) \leq \prod_{i=1}^4 {\rm dim}\, \T^i \eqno(D.6b) $$
where the reduced eigenvectors $\tf,\,\f$ are the (non-zero eigenvalue)
eigenvectors of the reduced problem
$$ \eqalign{ \;\;\;\;\;\; \;\;\;\;\;\;\;\;\;\;\;\;\;\;
 {\bM^{\a}(u_0)_{\b}}^{\g} \tf_{\g (\n)}^{\a} (u_0) & = E_{\n}^{\a}(u_0)
\tf_{\b(\n)}^{\a} (u_0) \cr
{\bM^{\a}(u_0)_{\b}}^{\g} & \equiv \sum_p W_{0p} (u_0 {)_{\b}}^{\a}
(Y_g (u_0) W_{p0} (u_0) )^{\g}  \cr} \eqno(D.7a)$$
$$ \eqalign{ \;\;\;\;\;\; \;\;\;\;\;\;\;\;\;\;\;\;\;\;
{M^{\a}(u_0)_{\b}}^{\g} \f_{\g (\n)}^{ \a} (u_0) & = E_{\n}^{\a}(u_0)
\f_{\b (\n)}^{ \a}  (u_0) \cr
{M^{\a}(u_0)_{\b}}^{\g} & \equiv \sum_q W_{q0}(u_0 {)_{\b}}^{\a}
(Y_g (u_0) W_{0q} (u_0))^{\g} \;\;\;\;. \cr} \eqno(D.7b)$$
To obtain this manifestly K-conjugation covariant form, we used the interchange
identity $ W_{q0 } W_{0p} = W_{0p} W_{q0} + \cO (k^{-2})$ in the $\ps$ problem.

Summing the series in (D.2), we obtain the high-level form of the
affine-Virasoro correlators
$$ \eqalign{ \bY_{\n}^{\a} (u,u_0) = Y_g^{\b}(u_0)  \left(
\d_{\b}^{\g} + {\tilde{P}^{ab} \over k} \right. &  \left[
\T_a^1  \T_b^2 \ln \left( {u \over u_0} \right) \right. \cr
 +\T_a^1  \T_b^3 & \left. \left.  \ln \left( {1-u \over 1-u_0} \right)
 \right]_{\b} {}^{\g}
 \right) \sqrt{E_{\n}^{\a} (u_0)}\,\tf_{\g(\n)}^{\a} (u_0)  + \cO (k^{-2}) \cr}
\eqno(D.8a)  $$
$$ \eqalign{ Y_{\n}^{\a} (u,u_0)  = Y_g^{\b}(u_0)  \left(
\d_{\b}^{\g} + {P ^{ab} \over k} \right. & \left[
\T_a^1  \T_b^2 \ln \left( {u \over u_0} \right) \right. \cr
 +\T_a^1  \T_b^3 & \left. \left. \ln \left( {1-u \over 1-u_0} \right)
 \right]_{\b} {}^{\g}
 \right) \sqrt{E_{\n}^{\a} (u_0)}\,\f_{\g(\n)}^{\a} (u_0)  + \cO (k^{-2})\;\;.
\cr}  \eqno(D.8b)  $$
Neglecting the constants $\sqrt{E_{\n}}\,(\tf,\f) $, this result is in
complete agreement with the high-level form of the vector ansatz in (10.8)

\end{document}